\documentclass[letterpaper,twocolumn,10pt]{article}
\usepackage{usenix,epsfig,endnotes}

\usepackage[total={6.5in,9in},paperheight=11in,paperwidth=8.5in]{geometry}
\usepackage{amsfonts,proof,qtree,amsmath,mathtools}
\usepackage{amsthm}
\usepackage{amssymb}
\usepackage{latexsym}
\usepackage{graphicx}
\usepackage{listings}
\usepackage{float}
\usepackage{cite}
\usepackage{multirow}
\usepackage[scaled]{helvet}
\usepackage[noend]{algorithmic}
\usepackage{mathrsfs}
\usepackage{mathpartir}
\usepackage{dsfont} 
\usepackage{stmaryrd}
\usepackage{url}
\usepackage{textcomp} 
\usepackage{mathpartir} 
\usepackage{caption}
\usepackage{multicol}
\usepackage{blindtext}
\usepackage{alltt} 
\usepackage{bbm}
\usepackage{alltt}
\usepackage{verbdef}
\usepackage{xspace}
\usepackage{verbatim}
\usepackage{enumitem}
\usepackage{lipsum}
\usepackage{wrapfig}
\usepackage{xcolor}
\usepackage{hyperref}
\hypersetup{linkcolor=black,citecolor=black,urlcolor=RubineRed}

\usepackage{enumitem}
\usepackage{pifont}
\usepackage[scaled]{helvet}
\usepackage{varwidth, shortcuts}
\usepackage{svg}
\usepackage{array}
\usepackage{enumitem}
\usepackage{combelow}
\usepackage{newunicodechar}
\makeatletter
\define@key{Gin}{page}{}
\makeatother

\definecolor[named]{ACMBlue}{cmyk}{1,0.1,0,0.1}
\definecolor[named]{ACMYellow}{cmyk}{0,0.16,1,0}
\definecolor[named]{ACMOrange}{cmyk}{0,0.42,1,0.01}
\definecolor[named]{ACMRed}{cmyk}{0,0.90,0.86,0}
\definecolor[named]{ACMLightBlue}{cmyk}{0.49,0.01,0,0}
\definecolor[named]{ACMGreen}{cmyk}{0.20,0,1,0.19}
\definecolor[named]{ACMPurple}{cmyk}{0.55,1,0,0.15}
\definecolor[named]{ACMDarkBlue}{cmyk}{1,0.58,0,0.21}

\hypersetup{colorlinks,
  linkcolor=ACMDarkBlue,
  citecolor=ACMPurple,
  urlcolor=ACMDarkBlue,
  filecolor=ACMDarkBlue}

\usepackage{booktabs}   %% For formal tables:

\definecolor{shadecolor}{gray}{1.00}
\definecolor{ddarkgray}{gray}{0.75}
\definecolor{darkgray}{gray}{0.30}
\definecolor{light-gray}{gray}{0.87}

\newcommand{\etc}{\emph{etc}}
\newcommand{\ie}{\emph{i.e.}\xspace}

\newcommand{\eg}{\emph{e.g.}\xspace}

\newcommand{\etal}{\emph{et~al.}\xspace}

\newcommand{\im}[1]{\mathsf{im}(#1)}

\newcommand{\cf}{\textit{cf.}\xspace}

\newcommand{\Iff}{\emph{iff}\xspace}

\theoremstyle{plain} 
{\unskip\nobreak\hskip 1em plus 1fil\nobreak$\square$
\parfillskip=0pt%
\endtrivlist}

\theoremstyle{definition}
\newtheorem{definition}{Definition}[section]

\newcommand{\is}[1]{{\textcolor{orange}{(Ilya: {#1})}}}

\newcommand{\True}{\mathsf{True}}

\newcommand{\set}[1]{\left\{{#1}\right\}}
\newcommand{\angled}[1]{\langle{#1}\rangle}

\newcommand{\args}{\mathit{args}}

\newcommand{\lab}{\ell}

\newcommand{\eqdef}{\triangleq}

\newcommand{\Nat}{\mathbb{N}}

\definecolor{pblue}{rgb}{0.13,0.13,1}
\definecolor{pgreen}{rgb}{0,0.5,0}
\definecolor{pred}{rgb}{0.9,0,0}
\definecolor{pgrey}{rgb}{0.46,0.45,0.48}

\definecolor{ckeyword}{HTML}{7F0055}
\definecolor{ccomment}{HTML}{3F7F5F}
\definecolor{cnumber}{HTML}{2A0099}

\lstdefinelanguage{Solidity}{
  keywords={typeof, transition, let, function, public, returns, external,
  contract, new, true, false, private, catch, function, return, null,
  throw, catch, switch, var, if, in, while, do, else, case, break,
  require},
  ndkeywords={bool, address, mapping, uint, bytes32, string, uint256},
  identifierstyle=\color{black},
  sensitive=false,
  comment=[l]{//},
  morecomment=[s]{/*}{*/},
  commentstyle=\color{ccomment}\ttfamily,
  string=[b]",
  morestring=[b]',
  showspaces=false,
  showtabs=false,
  breaklines=true,
  morekeywords={function, contract, returns, return},
  breakatwhitespace=true,
  lineskip=-0.6pt,
  basewidth={0.54em, 0.4em},%
  basicstyle=\scriptsize\ttfamily,
  keywordstyle={\color{ckeyword}\ttfamily\bfseries},
  ndkeywordstyle={\color{pblue}\ttfamily\bfseries},
  commentstyle={\color{ccomment}\itshape},
  stringstyle={\color{pred}\ttfamily},
  numberstyle={\scriptsize\color{cnumber}\sffamily},
  moredelim=[il][\textcolor{pgrey}]{$$},
  moredelim=[is][\textcolor{pgrey}]{\%\%}{\%\%}
}

\newcommand{\scode}[1]{\lstinline[language=Solidity,basicstyle=\small\ttfamily]{#1}}
\newcommand{\code}[1]{\scode{#1}}

\newcommand{\cstate}{\rho}

\newcommand{\tname}[1]{\textsc{#1}\xspace}
\newcommand{\elite}{\tname{EtherLite}}
\newcommand{\Id}{\mathit{id}}
\newcommand{\pc}{\mathit{pc}}
\newcommand{\bstate}{\sigma}
\newcommand{\state}{\delta}
\newcommand{\Bal}{\mathit{bal}}
\newcommand{\Code}{\mathit{code}}
\newcommand{\Sender}{\mathit{sender}}
\newcommand{\Value}{\mathit{value}}
\newcommand{\Data}{\mathit{data}}

\newcommand{\stepc}[1]{\xrightarrow[~~]{~#1~}}
\newcommand{\Lab}{\ell}

\newcommand{\Call}[1]{\mathsf{call}({#1})}

\newcommand{\Delegatecall}[1]{\mathsf{delegatecall}({#1})}

\newcommand{\Suicide}[1]{\mathsf{suicide}({#1})}

\newcommand{\Sstore}[1]{\mathsf{sstore}({#1})}
\newcommand{\Sload}[1]{\mathsf{sload}({#1})}

\newcommand{\rname}[1]{\textsc{#1}\xspace}
\newcommand{\last}[1]{\mathsf{last}(#1)}
\newcommand{\length}[1]{\mathsf{length}(#1)}

\newcommand{\ttrace}{\widehat{\tau}}
\newcommand{\Trace}{\tau}
\newcommand{\many}[1]{\overline{#1}}
\newcommand{\trace}[2]{\Trace_{#1}(#2)}

\newcommand{\predp}{\mathit{P}}
\newcommand{\predq}{\mathit{Q}}
\newcommand{\predr}{\mathit{R}}
\newcommand{\leaky}[1]{\mathit{leaky}_{#1}}
\newcommand{\locking}[1]{\mathit{locking}_{#1}}
\newcommand{\opn}[1]{{\tt {#1}}}

\setlist[itemize]{leftmargin=*}

\begin{document}

\author{
  {\rm Ivica Nikoli\'c}\\
  School of Computing, NUS\\
  Singapore \and
  {\rm Aashish Kolluri}\\
  School of Computing, NUS\\
  Singapore \and
  {\rm Ilya Sergey}\\
  University College London\\
  United Kingdom \and
  {\rm Prateek Saxena}\\
  School of Computing, NUS\\
  Singapore \and
  {\rm Aquinas Hobor}\\
  Yale-NUS College and School of Computing, NUS\\
  Singapore }

\setlist[enumerate]{leftmargin=0cm,itemindent=0.4cm,labelwidth=\itemindent,labelsep=0cm,align=left,partopsep=1pt,itemsep=-1ex,topsep=1pt}
\setlist[itemize]{topsep=0pt,itemsep=-1ex,partopsep=0ex,parsep=1ex,itemindent=0pt,leftmargin=8pt}

\title{\Large \bf 
  Finding The Greedy, Prodigal, and Suicidal Contracts at Scale
}

\maketitle

% Use the following at camera-ready time to suppress page numbers.
% Comment it out when you first submit the paper for review.
\thispagestyle{empty}

\begin{abstract}
Smart contracts---stateful executable objects hosted on blockchains
like Ethereum---carry billions of dollars worth of coins and cannot
be updated once deployed. We present a new systematic characterization
of a class of {\em trace vulnerabilities}, which result from analyzing
multiple invocations of a contract over its lifetime. 
We focus attention on three example properties of such trace
vulnerabilities: finding contracts that either lock funds
indefinitely, leak them carelessly to arbitrary users, or can be
killed by anyone.
We implemented \codename\footnote{\codename is available at \url{https://github.com/MAIAN-tool/MAIAN}.}
, the first tool for precisely specifying and
reasoning about trace properties, which employs inter-procedural
symbolic analysis and concrete validator for exhibiting real
%<<<<<<< HEAD
exploits. Our analysis of nearly one million contracts
flags \totalbuggy (\distinctbuggy distinct) contracts vulnerable, in
10 seconds per contract. On a subset of \validated contracts which we
sampled for concrete validation and manual analysis, we reproduce real
exploits at a true positive rate of \tpweight \%, yielding exploits
for \confirmedTP contracts. Our tool finds exploits for the infamous
Parity bug that indirectly locked $200$ million dollars worth in
Ether, which previous analyses failed to capture.

\end{abstract}

\section{Introduction}
\label{sec:introduction}

Cryptocurrencies feature a distributed protocol for a set of computers
to agree on the state of a public ledger called the blockchain.
Prototypically, these distributed ledgers map accounts or addresses
(the public half of a cryptographic key pair) with quantities of
virtual ``coins''. Miners, or the computing nodes, facilitate
recording the state of a payment network, encoding transactions that
transfer coins from one address to another. A significant number of
blockchain protocols now exist, and as of writing the market value of
the associated coins is over \$300 billion US, creating a lucrative
attack target.

{\em Smart contracts} extend the idea of a blockchain to a compute
platform for decentralized execution of general-purpose applications.
Contracts are programs that run on blockchains: their code and state
is stored on the ledger, and they can send and receive coins.  Smart
contracts have been popularized by the Ethereum blockchain.  Recently,
sophisticated applications of smart contracts have arisen, especially
in the area of token management due to the development of the ERC20
token standard.  This standard allows the uniform management of custom
tokens, enabling, \eg, decentralized exchanges and complex
wallets. Today, over a million smart contracts operate on the Ethereum
network, and this count is growing.

Smart contracts offer a particularly unique combination of security
challenges. Once deployed they cannot be upgraded or
patched,\footnote{Other than by ``hard forks'', which are essentially
  decisions of the community to change the protocol and are extremely
  rare.} unlike traditional consumer device software. Secondly, they are
written in a new ecosystem of languages and runtime environments, the
de facto standard for which is the Ethereum Virtual Machine and its
programming language called Solidity.
Contracts are relatively difficult to test, especially since their
runtimes allow them to interact with other smart contracts and
external off-chain services; they can be invoked repeatedly by
transactions from a large number of users.
Third, since coins on a blockchain often have significant value,
attackers are highly incentivized to find and exploit bugs in
contracts that process or hold them directly for profit.  The attack
on the DAO contract cost the Ethereum community \$60 million
US; and several more recent ones have had impact of a similar
scale~\cite{Parity}.

In this work, we present a systematic characterization of a class of
vulnerabilities that we call as {\em trace} vulnerabilities.  Unlike
many previous works that have applied static and dynamic analyses to
find bugs in contracts
automatically~\cite{Luu-al:CCS16,Oyente,Kalra-al:NDSS18,securify}, our
work focuses on detecting vulnerabilities across a long sequence of
invocations of a contract.  We label vulnerable contracts with three
categories --- greedy, prodigal, and suicidal --- which either lock
funds indefinitely, leak them to arbitrary users, or be susceptible to
by killed by any user. Our precisely defined properties capture many
well-known examples of known anecdotal
bugs~\cite{dao,Governmental,Parity}, but broadly cover a class of
examples that were not known in prior work or public reports. More
importantly, our characterization allows us to concretely check for
bugs by running the contract, which aids determining confirmed true
positives.

We build an analysis tool called \codename for finding these
vulnerabilities directly from the bytecode of Ethereum smart
contracts, without requiring source code access. In total, across the
three categories of vulnerabilities, \codename has been used to
analyze $970,898$ contracts live of the public Ethereum blockchain.
Our techniques are powerful enough to find the infamous Parity bug
that indirectly caused $200$ million dollars worth of Ether, which is
not found by previous analyses.  A total of \totalbuggy
(\distinctbuggy distinct) contracts are flagged as potentially buggy,
directly carry the equivalent of millions of dollars worth of Ether.
As in the case of the Parity bug, they may put a larger amount to
risk, since contracts interact with one another. For \validated
contracts we tried to concretely validate, \codename has found
over \confirmedTP confirmed vulnerabilities with \tpweight\% true
positive rate.  All vulnerabilities are uncovered on average within 10
seconds of analysis per contract.

\paragraph{Contributions}
We make the following contributions:

\begin{itemize}

\item We identify three classes of {\em trace} vulnerabilities, which
  can be captured as properties of a execution traces --- potentially
  infinite sequence of invocations of a contract.  Previous techniques
  and tools~\cite{Oyente} are not designed to find these bugs because
  they only model behavior for a single call to a contract.

\item We provide formal high-order properties to check which admit a
  mechanized symbolic analysis procedure for detection. We fully
  implement \codename, a tool for symbolic analysis of smart contract
  bytecode (without access to source code).

\item We test close to one million contracts, finding thousands of
  confirmed true positives within a few seconds of analysis time per
  contract. Testing trace properties with \codename is practical.

\end{itemize}

\section{Problem}

We define a new class of trace vulnerabilities, showing three specific
examples of properties that can be checked in this broader class. We
present our approach and tool to reason about the class of trace
vulnerabilities.

\subsection{Background on Smart Contracts}
\label{sec:background}
Smart contracts in Ethereum run on Ethereum Virtual Machine (EVM), a
stack-based execution runtime~\cite{Gavin-al:yellow-paper}. Different
source languages compile to the EVM semantics, the predominant of them
being Solidity~\cite{Solidity}. A smart contract embodies the
concept of an autonomous agent, identified by its program logic, its
identifying address, and its associated balance in Ether.
%
% The program is represented in its bytecode representation and it can
% access its contract balance programmatically via the variable
% \texttt{balance}.
%
Contracts, like other addresses, can receive Ether from external
agents storing it in their \code{balance} field; they can can also
send Ether to other addresses via transactions. A smart contract is
created by the {\em owner} who sends an initializing transaction,
which contains the contract bytecode and has no specified recipient.
Due to the persistent nature of the blockchain, once initialized, the
contract code cannot be updated. Contracts live perpetually unless
they are explicitly terminated by executing the \opn{SUICIDE}
bytecode instruction, after which they are no longer invocable or
called {\em dead}. When alive, contracts can be invoked many times.
Each invocation is triggered by sending a transaction to the contract
address, together with input data and a fee (known as
\emph{gas})~\cite{Gavin-al:yellow-paper}. The mining network executes
separate instances of the contract code and agrees on the outputs of
the invocation via the standard blockchain consensus protocol, \ie,
Nakamoto consensus~\cite{Nakamoto:08,Pirlea-Sergey:CPP18}. The result
of the computation is replicated via the blockchain and grants a
transaction fee to the miners as per block reward rates established
periodically.

The EVM allows contract functions to have local
state, while the contracts may have global variables stored on the
blockchain. Contracts can invoke other contracts via message calls;
outputs of these calls, considered to be a part of the same
transaction, are returned to the caller during the runtime.
Importantly, calls are also used to send Ether to other contracts and
non-contract addresses.
%
% Therefore, outputs of a contract invocation are blockchain state
% updates as well as transactions sending Ether to other addresses.
%
The \code{balance} of a contract can be read by anyone, but is only
updated via calls from other contracts and externally initiated
transactions.

Contracts can be executed repeatedly over their lifetime. A
transaction can run one {\em invocation} of the contract and an
execution \emph{trace} is a (possibly infinite) sequence of runs of a
contract recorded on the blockchain. Our work shows the importance of
reasoning about execution traces of contracts with a class of
vulnerabilities that has not been addressed in prior works, and
provides an automatic tool to detect these issues.

\subsection{Contracts with Trace Vulnerabilities}
\label{sec:bugs}

While trace vulnerabilities are a broader class, we our focus
attention on three example properties to check of contract
traces. Specifically, we flag contracts which (a) can be killed by
arbitrary addresses, (b) have no way to release Ether after a certain
execution state, and (c) release Ether to arbitrary addresses
carelessly.

Note that any characterization of bugs must be taken with a grain of
salt, since one can always argue that the exposed behavior embodies
intent --- as was debated in the case of the DAO bug~\cite{dao}. Our
characterization of vulnerabilities is based, in part, on anecdotal
incidents reported publicly~\cite{Governmental,parity-bug,dao}. To the
best of our knowledge, however, our characterization is the first to
precisely define checkable properties of such incidents and measure
their prevalence. Note that there are several valid reasons for
contracts for being killable, holding funds indefinitely under certain
conditions, or giving them out to addresses not known at the time of
deployment.
For instance, a common security best practice is that when under
attack, a contract should be killed and should return funds to a
trusted address, such as that of the owner.
Similarly, benign contracts such as bounties or games, often hold
funds for long periods of time (until a bounty is awarded) and release
them to addresses that are not known statically. 
Our characterization admits these benign behaviors and flags egregious
violations described next, for which we are unable to find justifiable
intent.

{\setlength{\belowcaptionskip}{-10pt}{
% Copyright 2017 Sergei Tikhomirov, MIT License
% https://github.com/s-tikhomirov/solidity-latex-highlighting/

%\usepackage{listings, xcolor}

\definecolor{verylightgray}{rgb}{.97,.97,.97}

\lstdefinelanguage{Solidity}{
	keywords=[1]{anonymous, assembly, assert, balance, break, call, callcode, case, catch, class, constant, continue, contract, debugger, default, delegatecall, delete, do, else, event, export, external, false, finally, for, function, gas, if, implements, import, in, indexed, instanceof, interface, internal, is, length, library, log0, log1, log2, log3, log4, memory, modifier, new, payable, pragma, private, protected, public, pure, push, require, return, returns, revert, selfdestruct, send, storage, struct, suicide, super, switch, then, this, throw, transfer, true, try, typeof, using, value, view, while, with, addmod, ecrecover, keccak256, mulmod, ripemd160, sha256, sha3}, % generic keywords including crypto operations
	keywordstyle=[1]\color{blue}\bfseries,
	keywords=[2]{address, bool, byte, bytes, bytes1, bytes2, bytes3, bytes4, bytes5, bytes6, bytes7, bytes8, bytes9, bytes10, bytes11, bytes12, bytes13, bytes14, bytes15, bytes16, bytes17, bytes18, bytes19, bytes20, bytes21, bytes22, bytes23, bytes24, bytes25, bytes26, bytes27, bytes28, bytes29, bytes30, bytes31, bytes32, enum, int, int8, int16, int24, int32, int40, int48, int56, int64, int72, int80, int88, int96, int104, int112, int120, int128, int136, int144, int152, int160, int168, int176, int184, int192, int200, int208, int216, int224, int232, int240, int248, int256, mapping, string, uint, uint8, uint16, uint24, uint32, uint40, uint48, uint56, uint64, uint72, uint80, uint88, uint96, uint104, uint112, uint120, uint128, uint136, uint144, uint152, uint160, uint168, uint176, uint184, uint192, uint200, uint208, uint216, uint224, uint232, uint240, uint248, uint256, var, void, ether, finney, szabo, wei, days, hours, minutes, seconds, weeks, years},	% types; money and time units
	keywordstyle=[2]\color{teal}\bfseries, 
	keywords=[3]{block, blockhash, coinbase, difficulty, gaslimit, number, timestamp, msg, data, gas, sender, sig, value, now, tx, gasprice, origin},	% environment variables
	keywordstyle=[3]\color{violet}\bfseries,
	identifierstyle=\color{black},
	sensitive=false,
	comment=[l]{//},
	morecomment=[s]{/*}{*/},
	commentstyle=\color{gray}\ttfamily,
	stringstyle=\color{red}\ttfamily,
	morestring=[b]',
	morestring=[b]"
}

\lstset{
	language=Solidity,
	backgroundcolor=\color{white},
	extendedchars=true,
	basicstyle=\scriptsize\ttfamily,
	%basicstyle=\fontsize{6}{9}\ttfamily,
	showspaces=false,
	numbers=left,
	numberstyle=\sffamily\tiny,
	numbersep=7pt,
	tabsize=2,
	breaklines=true,
	showtabs=false,
	captionpos=b,
	xleftmargin=2em,
	frame=single,
	framexleftmargin=0.5em,
	escapechar=\&
	% postbreak=\mbox{\textcolor{red}{$\hookrightarrow$}\space\space\space\space},
}

%\begin{multicols}{2}
%\blindtext
\begin{figure}[t!]
% \centering
%\begin{tabular}{cc}
%\begin{minipage}{0.50\linewidth}
\small
\begin{lstlisting}[mathescape=true,language=Solidity, frame=none, basicstyle=\fontsize{6}{9}\ttfamily]
function payout(address[] recipients,
                uint256[] amounts) {
 require(recipients.length==amounts.length);
 for (uint i = 0; i < recipients.length; i++) {
    /* ... */
    recipients[i].send(amounts[i]);
  }}
\end{lstlisting}
%\end{minipage}
%\end{tabular}
\captionof{figure}{\code{Bounty} contract; \code{payout} leaks Ether.
}
%\caption{\texttt{Buggy} contract.}
\label{fig:buggyPayout} 
%\end{figure*}
\end{figure}
%\end{multicols}
}}

%// transfer deposits funds to recipients
\begin{comment}
\is{Can we please make sure that the code snippets have no trailing
  lines translated in some ugly way, like what happens with the
  require-clause of the code above? Also, please, make sure that the
  numbering doesn't go to margins. Thanks.}

    	Payout(
      	msg.sender,
      	recipients[i],
      	i + 1,
      	amounts[i],
      	recipients[i].send(amounts[i])
    );}}

\end{comment}

\paragraph{Prodigal Contracts} 
Contracts often return funds to owners (when under attack), to
addresses that have sent Ether to it in past (\eg, in lotteries), or
to addresses that exhibit a specific solution (\eg, in bounties).
However, when a contract gives away Ether to an arbitrary address---
which is not an owner, has never deposited Ether in the contract, and
has provided no data that is difficult to fabricate by an arbitrary
observer---we deem this as a vulnerability. We are interested in
finding such contracts, which we call as \emph{prodigal}.

Consider the \code{Bounty} contract with code fragment given in Figure~\ref{fig:buggyPayout}. 
This contract collects Ether from different
sources and rewards  bounty to a selected set of recipients. 
% These recipients are selected by a set of privileged owners of the contract. 
In the contract, the function \code{payout} sends to a list of
recipients specified amounts of Ether.  It is clear from the function
definition that the recipients and the amounts are provided as inputs,
and anybody can call the function (\ie, the function does not have
restrictions on the sender).  The message sender of the transaction is
not checked for; the only check is on the size of lists.  Therefore,
any user can invoke this function with a list of recipients of her
choice, and completely drain its Ether.

The above contract requires a single function invocation to leak its
Ether. However, there are examples of contracts which need two or more
invocations (calls with specific arguments) to cause a leak. Examples
of such contracts are presented in Section~\ref{sec:eval}.

{\setlength{\belowcaptionskip}{-10pt}{
%\begin{multicols}{2}
%\blindtext

\begin{figure}[t!]
% \centering
%\begin{tabular}{cc}
%\begin{minipage}{0.50\linewidth}
\begin{lstlisting}[mathescape=true,language=Solidity, frame=none, basicstyle=\fontsize{6}{9}\ttfamily]
function initMultiowned(address[] _owners,
                        uint _required){
  if (m_numOwners > 0) throw;
  m_numOwners = _owners.length + 1;
  m_owners[1] = uint(msg.sender);
  m_ownerIndex[uint(msg.sender)] = 1;
  m_required = _required;
  /* ... */
}

function kill(address _to) {
    uint ownerIndex = m_ownerIndex[uint(msg.sender)];
    if (ownerIndex == 0) return;
    var pending = m_pending[sha3(msg.data)];
    if (pending.yetNeeded == 0) {
      pending.yetNeeded = m_required;
      pending.ownersDone = 0;
    }
    uint ownerIndexBit = 2**ownerIndex;
    if (pending.ownersDone &\&& ownerIndexBit == 0) {
      if (pending.yetNeeded <= 1) 
      	suicide(_to);
      else {
        pending.yetNeeded--;
        pending.ownersDone |= ownerIndexBit;
      }
    }
}


\end{lstlisting}
%\end{minipage}
%\end{tabular}
\captionof{figure}{Simplified fragment of \code{ParityWalletLibrary}
  contract, which can be killed.}
%\caption{\texttt{Buggy} contract.}
\label{fig:parity} 
%\end{figure*}
\end{figure}
%\end{multicols}
}}

\paragraph{Suicidal Contracts} 
A contract often enables a security fallback option of being killed by
its owner (or trusted addresses) in emergency situations like when being
drained of its Ether due to attacks, or when malfunctioning.  However,
if a contract can be killed by \emph{any} arbitrary account, which
would make it to execute the \opn{SUICIDE} instruction, we consider it
vulnerable and call it \emph{suicidal}.

The recent \emph{Parity} fiasco\cite{Parity} is a concrete example of
such type of a contract. A supposedly innocent Ethereum
user~\cite{Parityinnocent} killed a library contract on which the main
\emph{Parity} contract relies, thus rendering the latter
non-functional and locking all its Ether.  To understand the
\emph{suicidal} side of the library contract, focus on the shortened
code fragment of this contract given in Figure~\ref{fig:parity}.
%
%Eexplain the bug, we optimize the original code of the library contract and  mention the critical functions which got attacked. 
%One of the library contracts of \emph{Parity}  
%The contract which got killed is a library contract, shown in Figure~\ref{fig:parity}.  
%A supposedly innocent address~\cite{Parityinnocent} triggered a bug  in the contract  to become its owner, before killing it.
% 
%In order to explain the attack, we optimize the original code of the contract
%and  mention the critical functions which got attacked. 
To kill the contract, the user invokes two different functions: one to
set the ownership,\footnote{The bug would have been prevented has the
  function \texttt{initMultiowned} been properly initialized by the
  authors.} and one to actually kill the contract.
%
% \prateek{Can we really show the case of the two
%   invocations causing the kill? Isn't this the most interesting case,
%   why the hell are we talking in the air?}
%
%
That is, the user first calls \code{initMultiowned}, providing empty array for \code{_owners}, and zero for \code{_required}. 
This effectively means that the contract has no owners and that nobody has to agree to execute a specific contract function. 
%On line $2$, the variable \code{m_numowners} stores the number of owners of the contract.  When the
%attacker invoked the contract, it was not initialized i.e., the value of
%variable \code{m_numowners}  was $0$. 
%Hence, the attacker account could
%pass the condition on line $2$, letting the attacker account to  initialize
%the contract and making itself the owner of the contract on lines $3-5$.  Note
%that, the variable  \code{m_required} on line $6$ is set to $0$.
%
%
Then the user invokes the function \code{kill}. This function needs
\code{_required} number of owners to agree to kill the contract,
before the actual \code{suicide} command at line 22 is
executed. However, since in the previous call to
\code{initMultiowned}, the value of \code{_required} was set to
zero, \code{suicide} is executed, and thus the contract is killed.

%  One of its functions,
% \code{initMultiowned} shown in Figure X, assigns the addresses in the
% input array, as owners of the contract along with the message
% sender. However, the function mistakenly allows any arbitrary address
% to invoke it. Hence, an arbitrary address could send an array of
% addresses to become an owner of the contract.  
% In a real
% incident~\cite{Parityinnocent}, a supposedly innocent address
% triggered this bug to become the owner of the contract, before killing
% it using the \code{kill} function.

\paragraph{Greedy Contracts}
We refer to contracts that remain alive and lock Ether indefinitely,
allowing it be released under no conditions, as {\em greedy}.
%
\begin{comment}
\prateek{Why don't we take the example of the Parity bug here, first?
  The following text is put here to be approved by others}.
\end{comment}
%
%
In the example of the \code{Parity} contract, many other
\code{multisigWallet}-like contracts which held Ether, used functions
from the \code{Parity} library contract to release funds to their
users. After the \code{Parity} library contracts was killed, the wallet contracts could
no longer access the library, thus became greedy. This
vulnerability resulted in locking of \$200M US worth of
Ether indefinitely!

Greedy contracts can arise out of more direct errors as well.  The
most common such errors occur in contracts that accept Ether but
either completely lack instructions that send Ether out
(e.g. \code{send, call, transfer}), or such instructions are not
reachable. An example of contract that lacks commands that release
Ether, that has already locked Ether is given in
Figure~\ref{fig:examplelock}.

{\setlength{\belowcaptionskip}{-10pt}{

%\begin{multicols}{2}
%\blindtext
\begin{figure}[t!]
% \centering
%\begin{tabular}{cc}
%\begin{minipage}{0.50\linewidth}
\small
\begin{lstlisting}[mathescape=true,language=Solidity, frame=none, basicstyle=\fontsize{7}{9}\ttfamily]
contract AddressReg{
  address public owner;
  mapping (address=>bool) isVerifiedMap;
  function setOwner(address _owner){
    if (msg.sender==owner)
        owner = _owner;
  }
  function AddressReg(){ owner = msg.sender; }
  function verify(address addr){
    if (msg.sender==owner)
      isVerifiedMap[addr] = true;
  }
  function deverify(address addr){
    if (msg.sender==owner)
      isVerifiedMap[addr] = false;
    }
  function hasPhysicalAddress(address addr)
           constant returns(bool){
     return isVerifiedMap[addr];
  }
}
\end{lstlisting}
%\end{minipage}
%\end{tabular}
\captionof{figure}{\texttt{AddressReg} contract locks Ether.}
%\caption{\texttt{Buggy} contract.}
\label{fig:examplelock} 
%\end{figure*}
\end{figure}
%\end{multicols}
}}

\paragraph{Posthumous Contracts} 
%As a subset of the problem with locking Ether outlined above, let us consider the contract shown in Figure~\ref{fig:parity}. 
%
%On line $17$ the contract executed the \opn{SUICIDE} instruction that terminated its operation. 
When a contract is killed, its code and global variables are cleared
from the blockchain, thus preventing any further execution of its
code. However, all killed contracts continue to receive transactions.
Although such transactions can no longer invoke the code of the
contract, if Ether is sent along them, it is added to the contract
balance, and similarly to the above case, it is locked
indefinitely. Killed contract or contracts that do not contain any
code, but have non-zero Ether we call \emph{posthumous}.
%For instance, the aforementioned Parity library contract from Figure~\ref{fig:parity} is posthumous. 
It is the onus of the sender to check if the contract is alive before
sending Ether, and evidence shows that this is not always the
case. Because posthumous contracts require no further static analysis beyond
that for identifying suicidal contracts, we do not treat this as a separate
class of bugs. We merely list all posthumous contracts on the live Ethereum
blockchain we have found in Section~\ref{sec:eval}.

% These two instructions compile to a loop which runs over the entire
% storage of the contract to clear the global variables one by
% one. This process required 5057945 gas to execute whereas the
% maxiumum gas per transaction at that time was
% 4712388\cite{}. Although the ether was temporarily locked, this
% example provides clear insights into how ether can be locked inside
% a live contract.  Another example is sending ether to a suicided
% contract, the ether sent is indefinitely locked inside that account
% since there is no code to retrieve that ether once a contract is
% killed. According to our results, there are a humongous number of
% contracts that lock ether, on our blockchain.

%This type of bugs is characterised by inability of contracts to
%release Ether.  Such contracts \textbf{lock Ether}, i.e. they are
%able to receive Ether from certain (or all) accounts, but no single
%or multiple transactions from any account can make them to send their
%Ether.  An obvious downside of such contracts is that they lock Ether
%indefinitely, thus such Ether becomes unusable.

%
{\setlength{\belowcaptionskip}{-15pt}{
\begin{figure}[t]
\centering
\def\svgwidth{0.5\textwidth}
\input{tool_svg.pdf_tex}
\caption{\codename}
\label{fig:tool}
\end{figure}
}}

\subsection{Our Approach}
Each run of the contract, called an invocation, may exercise an
execution path in the contract code under a given input context.  Note
that prior works have considered bugs that are properties of
\emph{one} invocation, ignoring the chain of effects across a
\emph{trace} of
invocations~\cite{Luu-al:CCS16,Chen-al:SANER17,mythril,mueller-z3,securify,manticore}.

We develop a tool that uses systematic techniques to find contracts
that violate specific properties of traces. The violations are either:

(a) of \emph{safety} properties, asserting that there \emph{exists} a
trace from a specified blockchain state that causes the contract to
violate certain conditions; and

(b) of \emph{liveness} properties, asserting whether some actions
\emph{cannot} be taken in \emph{any} execution starting from a
specified blockchain state.

We formulate the three kinds of vulnerable contracts as these safety
and liveness trace properties in Section~\ref{sec:properties}. 
Our technique of finding vulnerabilities, implemented as a tool called
\codename and described in Section~\ref{sec:tool}, consists of two
major components: symbolic analysis and concrete validation.  The
symbolic analysis component takes contract bytecode and analysis
specifications as inputs.  The specifications include vulnerability
category to search for and depth of the search space, which further we
refer to as \emph{invocation depth}, along with a few other analysis
parameters we outline in Section~\ref{sec:tool}.  To develop our
symbolic analysis component, we implement a custom Ethereum Virtual
Machine, which facilitates symbolic execution of contract
bytecode~\cite{Oyente}.  With every contract candidate, our component
runs possible execution traces symbolically, until it finds a trace
which satisfies a set of predetermined properties. The input context
to every execution trace is a set of symbolic variables. Once a
contract is flagged, the component returns concrete values for these
variables.
Our final step is to run the contract concretely and validate the
result for true positives; this step is implemented by our concrete
validation component. The concrete validation component takes the
inputs generated by symbolic analysis component and checks the exploit
of the contract on a private fork of Ethereum blockchain. Essentially,
it is a testbed environment used to confirm the correctness of the
bugs.  As a result, at the end of validation the candidate contract is
determined as true or false positive, but the contract state on main
blockchain is not affected since no changes are committed to the
official Ethereum blockchain.

\section{Execution Model and Trace Properties}
\label{sec:properties}

A life cycle of a smart contract can be represented by a sequence of
the contract's states, which describe the values of the contract's
fields, as well as its balance, interleaved with instructions and
\emph{irreversible} actions it performs modifying the global context
of the blockchain, such transferring Ether or committing suicide.
One can consider a contract to be \emph{buggy} with respect to a
certain class of unwelcome high-level scenarios (\eg, ``leaking''
funds) if some of its finite execution traces fail to satisfy a
certain condition.
Trace properties characterised this way are traditionally qualified as
\emph{trace-safety} ones, meaning that ``during a final execution
nothing bad happens''.
Proving the absence of some other high-level bugs will, however,
require establishing a statement of a different kind, namely,
``something good must eventually happen''. Such properties are known
as \emph{liveness} ones and require reasoning about progress in
executions.
An example of such property would be an assertion that a contract can
always execute a finite number of steps in order to perform an action
of interest, such as tranferring money, in order to be considered
non-\emph{greedy}.

%  or executing a suicide command,
% \ie, to be non-greedy or mortal (as opposed to \emph{undead}).

In this section, we formally define the execution model of Ethereum
smart contracts, allowing one to pinpoint the vulnerabilities
characterised in Section~\ref{sec:bugs}.
The key idea of our bug-catching approach is to formulate the
erroneous behaviours as predicates of observed contract \emph{traces},
rather than individual configurations and instruction invocations,
occurring in the process of an execution.
By doing so, we are able to (a) capture the prodigal/suicidal
contracts via conditions that relate the unwelcome agents gaining, at
some point, access to a contract's funds or suicide functionality by
finding a way around a planned semantics, and (b) respond about
repeating behavioural patterns in the contract life cycles, allowing
us to detect greedy contracts.

% What do the high-level bugs from Section~\ref{sec:bugs} have in
% common?
% %
% We argue that the behavior deemed undesired in all those
% cases---greedy, wasteful, and suicidal contracts---can be precisely
% captured in a form of a \emph{trace} property of a contract execution.

\subsection{EVM Semantics and Execution Traces}
\label{sec:semantics}

We begin with defining cotnract execution traces by adopting a low-level
execution semantics of an EVM-like language in the form of \elite-like
calculus~\cite{Luu-al:CCS16}. \elite implements a small-step stack
machine, operating on top of a global configuration of the blockchain,
which used to retrieve contract codes and ascribe Ether balance to
accounts, as well as manipulations with the local contract
configuration.
%
% In our augmented version of \elite, in addition to the contract
% execution stack $S$ and the blockchain state will also store the
% latest \emph{message} $m$, sent by the most recent agent, which has
% invoked a contract.  
%
As customary in Ethereum, such agent is represented by its address
$\Id$, and might be a contract itself.
For the purpose of this work, we simplify the semantics of \elite by
eliding the executions resulting in exceptions, as reasoning about
such is orthogonal to the properties of interest. Therefore, the
configurations $\state$ of the \elite abstract machine are defined as follows:

{\small{
\[
\begin{array}{l@{\ \ \ \ }r@{\ \ }c@{\ \ }l}
  \text{Configuration} & \state & \eqdef & \angled{A, \bstate}
  \\[2pt]
  \text{Execution stack} & A &\eqdef & \angled{M, \Id, \pc, s, m} \cdot A ~~|~~ \epsilon
  \\[2pt]
  \text{Message} & m & \eqdef & \set{\Sender \mapsto \Id;~\Value : \Nat;~ \Data \mapsto \ldots}
  \\[2pt]
\text{Blockchain state} & \bstate & \eqdef & \many{\Id \mapsto \set{\Bal : \Nat;~\Code? \mapsto
                     M;~ \many{f? \mapsto v}}}
\end{array}
\]
}}

That is, a contract execution configuration consists of an activation
record stack $A$ and a blockchain context $\bstate$. An activation
record stack $A$ is a list of tuples $\angled{M, \Id, \pc, s, m}$,
where $\Id$ and $M$ are the address and the code of the contract
currently being executed, $\pc$ is a program counter pointing to the
next instruction to be executed, $s$ is a local operand stack, and $m$
is the last message used to invoke the contract execution. Among other
fields, $m$ stores the identity of the $\Sender$, the amount $\Value$
of the ether being transferred (represented as a natural number), as
well as auxiliary fields ($\Data$) used to provide additional
arguments for a contract call, which we will be omitting for the sake
of brevity.
Finally, a simplified context $\bstate$ of a blockchain is encoded as
a finite partial mapping from an account $\Id$ to its balance and
contract code $M$ and its mutable state, mapping the field names $f$
to the corresponding values,\footnote{For simplicity of presentation,
  we treat all contract state as \emph{persistent}, eliding operations
  with \emph{auxiliary} memory, such as
  \opn{MLOAD}/\opn{MSTORE}.} which both are optional (hence,
marked with ?) and are only present for contract-storing blockchain
records.
We will further refer to the union of a contract's fields entries
$\many{f \mapsto v}$ and its balance entry $\Bal \mapsto z$ as a
\emph{contract state}~$\cstate$.

{
\setlength{\abovecaptionskip}{-10pt}
\setlength{\belowcaptionskip}{-10pt}
{
\begin{figure}
{\footnotesize{
\begin{mathpar}
\inferrule*[Lab=SStore]
{
M[\pc] = \opn{SSTORE}
\and
\bstate' = \bstate[\Id][f \mapsto v]
}
{
\angled{\angled{M, \Id, \pc, f \cdot v \cdot s, m} \cdot A, \bstate}
  \stepc{\Sstore{f,~v}}
  \angled{\angled{M, \Id, \pc+1, s, m} \cdot A, \bstate'} }
\and
\inferrule*[Lab=SLoad]
{
M[\pc] = \opn{SLOAD}
\and
v = \bstate[\Id][f]
}
{
\angled{\angled{M, \Id, \pc, f \cdot s, m} \cdot A, \bstate}
  \stepc{\Sload{f,~v}}
\angled{\angled{M, \Id, \pc + 1, v \cdot s, m} \cdot A, \bstate}
}      
\and
\inferrule*[Lab=Call]
{
M[\pc] = \opn{CALL}
\and
\bstate[\Id][\Bal] \geq z
\\
s = \Id' \cdot z \cdot \args \cdot s'
\and
a = \angled{M, \Id, \pc + 1, s', m}
\\
m' = \set{\Sender \mapsto \Id; \Value \mapsto z; \Data \mapsto \args}
\and 
M' = \bstate[\Id'][\Code]
\\
\bstate' = \bstate[\Id][\Bal \mapsto \bstate[\Id][\Bal] - z]
\quad
\bstate'' = \bstate'[\Id'][\Bal \mapsto \bstate'[\Id'][\Bal] + z]
}
{
\angled{\angled{M, \Id, \pc, s, m} \cdot A, \bstate}
  \stepc{\Call{\Id',~m'}}
\angled{\angled{M', \Id', 0, \epsilon, m'} \cdot a \cdot A, \bstate''}
} 
\and
\inferrule*[Lab=SuicideNonEmptyStack]
{
M[\pc] = \opn{SUICIDE}
\and
s = \Id' \cdot s'
\and
a = \angled{M', \pc', s'', m'}
\\
\bstate' = \bstate[\Id'][\Bal \mapsto (\bstate[\Id'][\Bal] +
\bstate[\Id][\Bal])]
\quad
\bstate'' = \bstate'[\Id][\Bal \mapsto 0]
}
{
\angled{\angled{M, \Id, \pc, s, m} \cdot a \cdot A, \bstate}
\stepc{\Suicide{\Id'}}
  \angled{\angled{M', \Id', \pc', 1 \cdot s'', m'} \cdot A, \bstate''} }
\end{mathpar}
}}
  
\caption{Selected execution rules of \elite.}
\label{fig:opsem}
\end{figure}
}}

Figure~\ref{fig:opsem} presents selected rules for a smart contract
execution in \elite.\footnote{The remaining rules can be found in the
  work by Luu~\etal~\cite{Luu-al:CCS16}.} 
The rules for storing and loading values to/from a contract's field
$f$ are standard.
Upon calling another account, a rule \rname{Call} is executed, which
required the amount of Ether $z$ to be transferred to be not larger
than the contract $\Id$'s current balance, and changes the activation
record stack and the global blockchain context accordingly.
Finally, the rule \rname{SuicideNonEmptyStack} provides the semantics
for the $\opn{SUICIDE}$ instruction (for the case of a non-empty
activation record stack), in which case all funds of the terminated
contract $\Id$ are transferred to the caller's $\Id'$.
%
%
% \aashish{1) the funds could be transferred to a non contract account
%   too, this is there in several places 2)and I don't understand why
%   CONS in SUICIDECONS.}

An important addition we made to the semantics of \elite are execution
\emph{labels}, which allow to distinguish between specific transitions
being taken, as well as their parameters, and are defined as follows:

{\footnotesize{
\[
\begin{array}{rcl}
  \Lab  & \eqdef & \Sstore{f,~v} ~|~ \Sload{f,~v} ~|~ \Call{\Id,~m}
                   ~|~ \Suicide{\Id} ~|~ \ldots
\end{array}
\]
}}
\noindent
For instance, a transition label of the form $\Call{\Id,~m}$ captures
the fact that a currently running contract has transferred control to
another contract $\Id$, by sending it a message $m$, while the label
$\Suicide{\Id}$ would mean a suicide of the current contract, with
transfer of all of its funds to the account (a contract's or not)
$\Id$.
%
% \aashish{as mentioned earlier, id could be an address of an external
% non contract account. Similarly, in the suicide case.}

With the labelled operational semantics at hand, we can now provide a
definition of partial contract execution \emph{traces} as sequences of
interleaved contract states $\cstate_i$ and transition labels
$\Lab_j$ as follows:

\begin{definition}[Projected contract trace]
\label{def:ttrace}

A \emph{partial} projected trace $t = \ttrace_{\Id}(\bstate, m)$ of a
contract $\Id$ in an initial blockchain state $\bstate$ and an
incoming message $m$ is defined as a sequence
$[\angled{\cstate_0, \lab_0}, \ldots, \angled{\cstate_n, \lab_n}]$,
such that for every $i \in \set{0 \ldots n}$,
$\cstate_i = \bstate_i[\Id]|_{\Bal, \many{f}}$, where $\bstate_i$ is
the blockchain state at the $i^{\text{th}}$ occurrence of a
configuration of the form,
$\angled{\angled{\bullet, \Id, \bullet, \bullet, \bullet}, \bstate_i}$
in an execution sequence starting from the configuration
  $\angled{\angled {\bstate[\Id][\Code], \Id, 0, \epsilon, m} \cdot
    \epsilon, \bstate} $,
  and $\Lab_i$ is a label of an immediate next transition.
\end{definition}

In other words, $\ttrace_{\Id}(\bstate, m)$ captures the states of a
contract $\Id$, interleaved with the transitions taken ``on its
behalf'' and represented by the corresponding labels, starting from the
initial blockchain $\bstate$ and triggered by the message $m$.
The notation $\bstate[\Id]|_{\Bal, \many{f}}$ stands for a projection
to the corresponding components of the contract entry in~$\bstate$.
States and transitions of contracts \emph{other} than $\Id$ and
involved into the same execution are, thus, ignored.

Given a (partial) projected trace $\ttrace_{\Id}(\bstate, m)$, we say
that it is \emph{complete}, if it corresponds to an execution, whose
last configuration is $\angled{\epsilon, \bstate'}$ for some
$\bstate'$.
The following definition captures the behaviors of multiple subsequent
transactions with respect to a contract of interest.

\begin{definition}[Multi-transactional contract trace]
\label{def:trace}
A contract trace $t = \trace{\Id}{\bstate, \many{m_i}}$, for a
sequence of messages $\many{m_i} = m_0, \ldots, m_n$, is a
concatenation of single-transaction traces
$\ttrace_{\Id}(\bstate_i, m_i)$, where $\bstate_0 = \bstate$,
$\bstate_{i+1}$ is a blockchain state at the end of an execution
starting from a configuration
$\angled{\angled {\bstate[\Id][\Code], \Id, 0, \epsilon, m_i} \cdot
  \epsilon, \bstate_i}$,
and all traces $\ttrace_{\Id}(\bstate_i, m_i)$ are complete for
$i \in \set{0, \ldots, n - 1}$.
\end{definition}

As stated, the definition does \emph{not} require a trace
to end with a complete execution at the last transaction.
For convenience, we will refer to 
%
% specific elements
% $\angled{\cstate_i, \Lab_i}$ of a trace $t$ as $t_i$, using the
% notation $t_i^1$ for $\cstate_i$ and $t_i^2$ for $\Lab_i$,
% correspondingly. We will also denote 
%
the last element of a trace $t$ by $\last{t}$ and to its length as
$\length{t}$.

\subsection{Characterising Safety Violations}
\label{sec:trace-safety-prop}

The notion of contract traces allows us to formally capture the
definitions of buggy behaviors, described previously in
Section~\ref{sec:bugs}. First, we turn our attention to the
prodigal/suicidal contracts, which can be uniformly captured by the
following higher-order trace predicate.

\begin{definition}[Leaky contracts]
\label{def:leaky}
A contract with an address $\Id$ is considered to be \emph{leaky} with
respect to predicates $\predp$, $\predr$ and $\predq$, and a
blockchain state~$\bstate$ (denoted
$\leaky{\predp, \predr, \predq}(\Id, \bstate)$) \Iff \emph{there
  exists} a sequence of messages $\many{m_i}$, such that for a trace
$t = \trace{\Id}{\bstate, \many{m_i}}$:

\begin{enumerate}
\item the \emph{precondition} $\predp(\bstate[\Id][\Code], t_0, m_0)$
  holds,
\item the \emph{side condition} $\predr(t_i, m_0)$ holds for all
  $i < \length{t}$,
\item the \emph{postcondition} $\predq(t_n, m_0)$ holds for
  $t_n = \last{t}$.
\end{enumerate}
  
\end{definition}

Definition~\ref{def:leaky} of leaky contracts is \emph{relative} with
respect to a current state of a blockchain: a contract that is
currently leaky may stop being such in the future.
Also, notice that the ``triggering'' initial message $m_0$ serves as
an argument for all three parameter predicates.
%
% \footnote{Possible
%   trade-offs of our definition are discussion in
%   Section~\ref{sec:discussion}.}
%
We will now show how two behaviors observed earlier can be encoded via
specific choices of $\predp$, $\predr$, and $\predq$.\footnote{In most
  of the cases, it is sufficient to take $R \eqdef \True$, but in
  Section~\ref{sec:related} we hint certain properties that require a
  non-trivial side condition.}

\paragraph{Prodigal contracts}
A contract is considered prodigal if it sends Ether, immediately or
after a series of transitions (possibly spanning multiple
transactions), to an arbitrary sender. This intuition can be
encoded via the following choice of $\predp$, $\predr$, and $\predq$
for Definition~\ref{def:leaky}:

{\footnotesize{
\[
\begin{array}{lcl@{\ \ \ }l}
  \predp(M, \angled{\cstate, \Lab}, m)  & \eqdef 
  &&
    m[\Sender] \notin \im{\cstate} \wedge 
    m[\Value] = 0
\\[2pt]
\predr(\angled{\cstate, \Lab}, m) & \eqdef && \True
\\[2pt]
\predq(\angled{\cstate, \Lab}, m) & \eqdef &&
\Lab = \Call{m[\Sender], m'} \wedge m'[\Value] > 0 
% \\[2pt]
% &&\vee& \Lab = \Call{m[\Sender], m'} \wedge m'[\Value] > 0 
\\[2pt]
&&\vee& \Lab = \Delegatecall{m[\Sender]}
\\[2pt]
&& \vee&\Lab = \Suicide{m[\Sender]}
\end{array}
\]
}}
%
% \aashish{I suggest including the EtherLite rules of Delegatecall in
% Fig 5 or explain the semantics of delegate call like we have
% explained for call} \is{FWIK, EthereLite does not have a rule for
% delegate call, and I'm a bit reluctant to invent it and add it now.}
%
%
According to the instantiation of the parameter predicates above, 
a prodigal contract is exposed by a trace that is triggered by a
message $m$, whose sender does \emph{not} appear in the contract's
state ($m[\Sender] \notin \im{\cstate}$), \ie, it is not the owner,
and the Ether payload of $m$ is zero. 
To expose the erroneous behavior of the contract, the postcondition
checks that the transition of a contract is such that it transfer
funds or control (\ie, corresponds to \opn{CALL}, \opn{DELEGATECALL}
or \opn{SUICIDE} instructions~\cite{Gavin-al:yellow-paper}) with the
recipient being the sender of the initial message. In the case of
sending funds via \opn{CALL}
 % or \opn{SEND} 
we also check that the
amount being transferred is non zero.
In other words, the initial caller $m[\Sender]$, unknown to the
contract, got himself some funds without any monetary contribution!
In principle, we could ensure minimality of a trace, subject to the
property, by imposing a non-trivial side condition $\predr$, although
this does not affect the class of contracts exposed by this
definition.

\paragraph{Suicidal contracts}
A definition of a suicidal contract is very similar to the one of a
prodigal contract. It is delivered by the following choice of
predicates:

{\footnotesize{
\[
\begin{array}{lcl}
  \predp(M, \angled{\cstate, \Lab}, m)  & \eqdef 
  &
    \opn{SUICIDE} \in M \wedge m[\Sender] \notin \im{\cstate}
\\[2pt]
\predr(\angled{\cstate, \Lab}, m) & \eqdef & \True
\\[2pt]
\predq(\angled{\cstate, \Lab}, m) & \eqdef & \Lab = \Suicide{m[\Sender]}
\end{array}
\]
}}

That is, a contract is suicidal if its code $M$ contains the
$\opn{SUICIDE}$ instruction and the corresponding transition can be
triggered by a message sender, that does not appear in the contract's
state at the moment of receiving the message, \ie, at the initial
moment $m[\Sender] \notin \im{\cstate}$.

\subsection{Characterising Liveness Violations}
\label{sec:liveness-properties}

A contract is considered \emph{locking} at a certain blockchain state
$\bstate$, if at any execution originating from $\bstate$ prohibits
certain transitions to be taken. Since disproving liveness properties
of this kind with a finite counterexample is impossible in general, we
formulate our definition as an \emph{under-approximation} of the
property of interest, considering only final traces up to a certain
length:

\begin{definition}[Locking contracts]
\label{def:locking}
A contract with an address $\Id$ is considered to be \emph{locking}
with respect to predicates $\predp$ and $\predr$, the transaction
number $k$, and a blockchain state~$\bstate$ (denoted
$\locking{\predp, \predr, k}(\Id, \bstate)$) \Iff \emph{for all}
sequences of messages $\many{m_i}$ of length less or equal than $k$,
the corresponding trace $t = \trace{\Id}{\bstate, \many{m_i}}$
satisfies:

\begin{enumerate}
\item the \emph{precondition} $\predp(\bstate[\Id][\Code], t_0, m_0)$,
\item the \emph{side condition} $\predr(t_i, m_0)$ for all
  $i \leq \length{t}$.
\end{enumerate}
  
\end{definition}

Notice that, unlike Definition~\ref{def:leaky}, this Definition does
not require a postcondition, as it is designed to under-approximate
potentially infinite traces, up to a certain length $k$,\footnote{We
  discuss viable choices of $k$ in Section~\ref{sec:eval}.} so the
``final state'' is irrelevant.

\paragraph{Greedy contracts}
In order to specify a property asserting that in an interaction with
up to $k$ transactions, a contract does not allow to release its
funds, we instantiate the predicates from Definition~\ref{def:locking}
as follows:

{\footnotesize{
\[
\begin{array}{l@{\ \ }c@{\ \ }l}
  \predp(M, \angled{\cstate, \Lab}, m)  & \eqdef 
  &\cstate[\Bal] > 0
\\[2pt]
\predr(\angled{\cstate, \Lab}, m) & \eqdef &

\neg\left(\!\! 
\begin{array}{l@{\ \ }l}
&
\Lab = \Call{m[\Sender], m'} \wedge m'[\Value] > 0 
\\[2pt]
\vee& \Lab = \Delegatecall{m[\Sender]}
\\[2pt]
\vee&\Lab = \Suicide{m[\Sender]}
\end{array}
\!\!\right)
\end{array}
\]
}}

Intuitively, the definition of a greedy contract is dual to the notion
of a prodigal one, as witnessed by the above formulation: at any trace
starting from an initial state, where the contract holds a non-zero
balance, no transition transferring the corresponding funds (\ie,
matched by the side condition $\predr$) can be taken, no matter what
is the $\Sender$'s identity. That is, this definition covers the case
of contract's \emph{owner} as well: \emph{no one} can withdraw any
funds from the contract.

\section{The Algorithm and the Tool}
\label{sec:tool}

\codename is a symbolic analyzer for smart contract execution traces,
for the properties defined in Section~\ref{sec:properties}.
It operates by taking as input a contract in its bytecode form and a
concrete starting block value from the Ethereum blockchain as the
input context, flagging contracts that are outlined in
Section~\ref{sec:bugs}. 
When reasoning about contract traces, \codename follows the \elite
rules, described in Section~\ref{sec:semantics}, executing them
symbolically. 
During the execution, which starts from a contract state satisfying
the precondition of property of interest (\cf
Definitions~\ref{def:leaky} and~\ref{def:locking}), it checks if there
exists an execution trace which violates the property and a set of
candidate values for input transactions that trigger the property
violation.
For the sake of tractability of the analysis, it does not keep track
of the entire blockchain context~$\bstate$ (including the state of
other contracts), treating only the contract's transaction inputs and
certain block parameters as symbolic.
To reduce the number of false positives and confirm concrete exploits
for vulnerabilities, \codename calls its {\em concrete validation}
routine, which we outline in Section~\ref{sec:validation}.

\subsection{Symbolic Analysis}
\label{sec:symbolic}

Our work concerns finding properties of traces that involve multiple
invocations of a contract. We leverage static symbolic analysis to
perform this step in a way that allows reasoning across contract calls
and across multiple blocks. We start our analysis given a contract
bytecode and a starting concrete context capturing values of the
blockchain. 
\codename reasons about values read from input transaction fields and
block parameters\footnote{Those being \opn{CALLVALUE}, \opn{CALLER},
  \opn{NUMBER}, \opn{TIMESTAMP}, \opn{BLOCKHASH}, \opn{BALANCE},
  \opn{ADDRESS}, and \opn{ORIGIN}.} in a symbolic way---specifically,
it denotes the set of all concrete values that the input variable can
take as a symbolic variable. It then symbolically interprets the
relationship of other variables computed in the contract as a symbolic
expression over symbolic variables. For instance, the code \code{y :=
  x + 4} results in a symbolic value for \code{y} if \code{x} is a
symbolic expression; otherwise it is executed as concrete
value. Conceptually, one can imagine the analysis as maintaining two
memories mapping variables to values: one is a symbolic memory mapping
variables to their symbolic expressions, the other mapping variables
to their concrete values.
%
% A number of excellent introductions to such symbolic execution
% techniques are available~\cite{klee,all-i-wanted}.

\paragraph{Execution Path Search}
The symbolic interpretation searches the space of all execution paths
in a trace with a depth-first search. The search is a best effort to
increase coverage and find property violating traces. Our goal is
neither to be sound, \ie, search all possible paths at the expense of
false positives, nor to be provably complete, \ie, have only true
positives at the expense of
coverage~\cite{Godefroid:2011}. From a practical perspective,
we make design choices that strike a balance between these two goals.

The symbolic execution starts from the entry point of the contract,
and considers all functions which can be invoked externally as an
entry point. More precisely, the symbolic execution starts at the first instruction in the bytecode, 
proceeding sequentially until the execution path ends in terminating instruction. 
Such instruction can be valid (\eg, \opn{STOP, RETURN}), in which case we assume to have reached the end of some contract function, and thus restart the symbolic execution again from the first bytecode instruction to simulate the next function call.
On the other hand, the terminating instruction can be invalid (\eg, non-existing instruction code or invalid jump destination), in which case we terminate the search down this path and backtrack in the depth-first search procedure to try another path.
When execution reaches a branch, \codename
concretely evaluates the branch condition if all the variables used in
the conditional expression are concrete. This uniquely determines the
direction for continuing the symbolic execution. If the condition
involves a symbolic expression, \codename queries an external SMT
solver to check for the satisfiability of the symbolic conditional
expression as well as its negation. Here, if the symbolic conditional
expression as well as its negation are satisfiable, both branches are
visited in the depth-first search; otherwise, only the satisfiable
branch is explored in the depth first search. On occasions, the
satisfiability of the expression cannot be decided in a pre-defined
timeout used by our tool; in such case, we terminate the search down
this path and backtrack in the depth-first search procedure to try
another path. We maintain a symbolic path constraint which captures
the conditions necessary to execute the path being analyzed in a
standard way. \codename implements support for 121 out of the 133 bytecode
instructions in Ethereum's stack-based low-level language.

At a call instruction, control follows transfer to the target. If the
target of the transfer is a symbolic expression, \codename backtracks
in its depth-first search. Calls outside a contract, however, are not
simulated and returns are marked symbolic. Therefore, \codename
depth-first search is inter-procedural, but not inter-contract.

\paragraph{Handling data accesses}
The memory mappings, both symbolic and concrete, record all the
contract memory as well blockchain storage.  During the symbolic
interpretation, when a global or blockchain storage is accessed for
the first time on a path, its concrete value is read from the main Ethereum blockchain  into local
mappings. This ensures that subsequent reads or writes are kept local to
the path being presently explored.  

The EVM machine supports a flat byte-addressable memory,
and each address has a bit-width of 256 bits. The accesses are in
32-byte sized words which \codename encodes as bit-vector constraints
to the SMT solver. Due to unavailability of source code, \codename does
not have any prior information about higher-level datatypes in
the memory. All types default to 256-bit integers in the encoding used
by \codename. Furthermore, \codename attempts to recover more advanced
types such as dynamic arrays by using the following heuristic:
%to speed up its SMT query resolution.  
%It uses the following heuristic: 
if a symbolic variable,
say $x$, is used in constant arithmetic to create an expression (say
$x + 4$) that loads from memory (as an argument to the
\opn{CALLDATALOAD} instruction), then it detects such an access as a
dynamic memory array access. Here, \codename uses the SMT solver to
generate $k$ concrete values for the symbolic expression, making the
optimistic assumption that the size of the array to be an integer in
the range $[0,k]$. The parameter $k$ is configurable, and
defaults to $2$.
Apart from this case, whenever accesses in the memory involve a
symbolic address, \codename makes no attempt at alias analysis and
simply terminates the path being search and backtracks in its
depth-first search.

\paragraph{Handling non-deterministic inputs}
Contracts have several sources of non-deterministic inputs such as the
block timestamp, \etc. While these are treated as symbolic, these are
not exactly under the control of the external users. \codename does
not use their concrete values as it needs to reason about invocations
of the contract across multiple invocations, \ie, at different blocks.

% \begin{comment}
% Another source of non-determinism
% are calls to off-chain code, typically performed via the
% \texttt{Oraclize} API. \prateek{What do we do for Oraclize precisely,
%   when there are concrete inputs and when there are symbolic inputs?}
% \end{comment}

\paragraph{Flagging Violations}
Finally, when the depth-first search in the space of the contract
execution reaches a state where the desired property is violated, it
flags the contract as a buggy {\em candidate}. The symbolic path
constraint, along with the necessary property conditions, are asserted
for satisfiability to the SMT solver. We
use~\tname{Z3}~\cite{deMoura-Bjorner:TACAS08} as our solver, which
provides concrete values that make the input formula satisfiable.  We
use these values as the concrete data for our symbolic inputs,
including the symbolic transaction data.
%
% \prateek {We need to explain how to check the conditions here, perhaps
%   with an example.}
%
\paragraph{Bounding the path search space}
\codename takes the following steps to bound the search in the
(potentially infinite) path space. First, the call depth is limited to
the constant called \code{max_call_depth}, which defaults to 3 but can
be configured for empirical tests. Second, we limit the total number
of jumps or control transfers on one path explored to a configurable
constant \code{max_cfg_nodes}, default set to $60$. This is
necessary to avoid being stuck in loops, for instance. Third, we set a
\code{timeout} of $10$ seconds per call to our SMT solver. 
% \begin{comment}
% Fourth, \codename simulates the usage of \texttt{gas} for each instruction it
% simulates faithfully; therefore, certain execution paths terminate as
% a result of insufficient gas and the search back-tracks at these
% points.  
% \end{comment}
Lastly, the total time spent on a contract is limited to configurable
constant \code{max_analysis_time}, default set to $300$ seconds.

\paragraph{Pruning}
To speed up the state search, we implement pruning with
memorization. Whenever the search encounters that the particular
configuration (\ie, contract storage, memory, and stack) has been seen
before, it does not further explore that part of the path space.

\subsection{Concrete Validation}
\label{sec:validation}

In the concrete validation step, \codename creates a private fork of
the original Ethereum blockchain with the last block as the input
context. It then runs the contract with the concrete values of the
transactions generated by the symbolic analysis to check if the
property holds in the concrete execution. If the concrete execution
fails to exhibit a violation of the trace property, we mark the
contract as a {\em false positive}; otherwise, the contract is marked
as a {\em true positive}.
To implement the validating framework, we added a new functionality to
the official go-ethereum package~\cite{goEthereum} which allows us to
fork the Ethereum main chain at a block height of our choice.
Once we fork the main chain, we mine on that fork without connecting to any peers on the Ethereum network, 
%, bythis, we are able to mine our own blocks without letting  the network know about them.  Consequently, 
and thus we are able to mine our own transactions  without
committing them to the main chain. 
%One has to observe that 
%Note, we generate our own
%Ether, required for validating purposes. For instance, any transaction
%submitted to the blockchain requires gas  to be mined(executed) and in order
%to provide that gas, the sender's account needs to have Ether  balance. 
% During exploitation of a contract, 
%On our private chain, the initial contract state is its state at the block
%height where we have forked the main chain. 
%The contract's state may change with every
%invocation of the exploit. However,  once the validation process terminates,
%it leaves the contract's  state unchanged on the main chain since the mined
%blocks on our fork are never shown to the network.

\paragraph{Prodigal Contracts} The validation framework checks if a
contract indeed leaks Ether by sending to it the transactions with
inputs provided by the symbolic analysis engine. The transactions are
sent by one of our accounts created previously.
%(message sender) which does
%not mine, submits transactions to the network and our miner account mines
%these transactions . 
Once the transactions are executed, the validation framework checks
whether the contract has sent Ether to our account.
%the message sender to confirm whether  the candidate is a True Positive.  
If a verifying contract does not have Ether, our framework first sends
Ether to the contract and only then runs the exploit.

%If the contract does not exist on main chain then the main chain can be forked at a
%block height where the contract exists, before exploiting it.

\paragraph{Suicidal Contracts} %In this case, 
In a similar fashion, the framework checks if a
contract can be killed after executing the transactions provided by the symbolic analysis engine 
on the forked chain. 
%performing the exploit on forked chain by a similar procedure described in the case of Prodigal contracts. 
Note, once a contract is killed, %its  storage fields are cleared on the blockchain and 
its bytecode is reset to \texttt{'0x'}. % and  all its balance is sent to a specific address which is decided by the contracts logic. 
Our framework uses precisely this test to confirm the correctness of the exploit. 
%checks if  for the bytecode field after performing an exploit. 
%If a contract doesn't exist on the blockchain at that block height then fork can be started at a
%block height where  the contract exists.

\paragraph{Greedy Contracts} A strategy similar to the above two
cannot be used to validate the exploits on contracts that lock
Ether. However, during the bug finding process, our symbolic execution
engine checks firsthand whether a contract accepts Ether.  
The validation framework can, thus, check if a contract is true
positive by confirming that it accepts Ether and does not have
\texttt{CALL}, \texttt{DELEGATECALL}, or
\texttt{SUICIDE} opcodes in its bytecode. In Section~\ref{sec:eval} we
give examples of such contracts.

\vspace{-5pt}

\section{Evaluation}
\label{sec:eval}

% We have only analyzed bytecode of the smart contracts. In that way, we
% are able to omit all the disadvantages of analyzing just the source
% code. We analyze contracts pertaining to the first 4.8 million
% blocks(4800000) blocks. In total, there are \emph{x} contracts, with y
% percentage of them having source on etherscan. Although a source could
% help understand the contract behaviour, it is laborious and a human
% may not always go through all the possibilities of bugs. Also, many
% times a developer of a smart contract may not want to publish source
% code at all on the blockchain. In such cases, an external user may
% verify the smart contract himself hence, mitigating the need for
% trusting the developers. In fact, the tool can be built into the
% ethereum system inorder to prevent upload of malicious contracts on
% the blockchain in the first place. The parity bug wouldn't have
% occured if the developers had tested their contract using our tool
% before uploading it on the blockchain.
% 
% \aashish{Write about the increase in the number of contract on block chain, the value of contracts}\\
% \aashish{Compare the number of contracts with source code and without on the blockchain and also their trends.}\\
% \aashish{Show why analyzing smart contracts with only bytecode is necessary by comparing their total value
% with the ones with source code.}\\
% 
% \aashish{Write about the dataset and their total loc and from where do we obtain it.} 
% 

% \paragraph{Dataset} 
We analyzed $970,898$ smart contracts, obtained by downloading the
Ethereum blockchain from the first block utill block number
$4,799,998$, which is the last block as of December $26$,
2017. Ethereum blockchain has only contract bytecodes. To obtain the
original (Solidity) source codes, we refer to the Etherscan
service~\cite{etherscanSourceCodes} and obtain source for $9,825$
contracts. Only around $1\%$ of the contracts have source code,
highlighting the utility of \codename as a bytecode analyzer.

% On the original blockchain, only the smart contract bytecode is
% available. To obtain the original source code for these contracts, we
% used the official EtherScan website. Of nearly 1m contract with bytecode,
% we find source code for $12,434$. Less than $2\%$ of the contracts
% have source code, which is a motivation for building a tool that 
% analyzes bytecode of contracts.

% There are number of reverse engineering tools to
% reverse engineer to source~\cite{porosity}. Our early attempts at
% using Porosity, one of the most widely used source-recovery tool,
% revealed an accuracy of below $50\%$; hence, we do not use or make
% deductions based on source recovery. Our experiment is reported in
% Table 1.

Recall that our concrete validation component can analyze a contract
from a particular block height where the contract is alive (\ie,
initialized, but not killed). To simplify our validation process for a
large number of contracts flagged by the symbolic analysis component,
we perform our concrete validation at block height of $4,499,451$,
further denoted as \texttt{BH}.  At this block height, we find that
most of the flagged contracts are alive, including the Parity library
contract~\cite{Parity} that our tool successfully finds. This contract
was killed at a block height of $4,501,969$. All contracts existing on
blockchain at a block height of $4,499,451$ are tested, but only
contracts that are alive at \texttt{BH} are concretely
validated.\footnote{We also concretely validate the flagged candidates
  which were killed before \texttt{BH} as well.}

\paragraph{Experimental Setup and Performance} 
\codename supports parallel analysis of contracts, and scales linearly
in the number of available cores.  We run it on a Linux box, with
64-bit Ubuntu 16.04.3 LTS, 64GB RAM and 40 CPUs Intel(R) Xeon(R)
E5-2680 v2@2.80GHz.  In most of our experiments we run the tool on
$32$ cores.
On average, \codename requires around 10.0 seconds to analyze a
contract for the three aforementioned bugs: 5.5 seconds to check if a
contract is prodigal, 3.2 seconds for suicidal, and $1.3$ seconds for
greedy.

\paragraph{Contract Characteristics}
The number of contracts has increased tenfold from Dec, $2016$ to Dec,
$2017$ and $176$-fold since Dec, $2015$.  However, the distribution of
Ether balance across contracts follows a skewed distribution.
% as shown in Figure~\ref{fig:baldist}
%
Less than $1\%$ of the contracts have more than $99\%$ of the Ether in
the ecosystem. This suggests that a vulnerability in any one of these
high-profile contracts can affect a large fraction of the entire Ether
balance. Note that contracts interact with each other, therefore, a
vulnerability in one contract may affect many others holding Ether, as
demonstrated by the recent infamous Parity library which was used by
wallet contracts with $\$200$ million US worth of Ether~\cite{Parity}.

% \paragraph{Contract Characteristics}.The number of contracts has increased 10 fold from
% Dec, 2016 to Dec 2017, and $176x$ since Dec 2015.

% - The distribution of balance across contracts follows a skewed
% distribution, as shown in Figure \ref{fig:baldist}.  Less than $1\%$ of the contracts
% have $99\%$ of the Ether in the ecosystem.

% % - Similarity of contracts. We clustered contracts based on edit
% % distance / similarity directly on bytecode. We find about 1400
% % different clusters. This is shown in Figure B.

% - The distribution of bytcode size is plotted in Figure \ref{fig:sizedist}. The highest
% and lowest sizes are $2$ and $23,735$ bytes, respectively. Average is $~1,073$ bytes. 
% Similarly, number of functions in contracts is plotted in Figure \ref{fig:funcdist}. 

% - Contract usage, \ie, the total number of transactions on the
% blockchain, varies widely on the contract. The total number of
% externally invoked transactions is $112,000,000$ involving only
% contracts (not direct exchange of Ether).

% - The average contract balance for the set of contracts that have
% received 1 or more transactions is $5x$ higher than that of all
% contracts. There is a direct correlation between how much Ether
% contracts hold and their usage.

% \begin{figure}[t]
% %  \fbox{
% % \begin{varwidth}{
% % %\dimexpr\textwidth+2\fboxsep-2\fboxrule\relax
% % }
% \centering
% %hereivica
%   \includegraphics[width=0.43\textwidth]{contractstats/baldistnomag.png}  
% %\end{varwidth}}
%   \caption{Distribution of Ether balance among contracts.}
% \label{fig:baldist} 
% \end{figure}
{\setlength{\belowcaptionskip}{-10pt}{
\begin{table}[t]
\centering
\footnotesize
% \begin{tabular}{|p{1.2 cm}||p{1.2 cm}|p{1.5 cm}|p{1.5 cm}|p{1.5 cm}|p{0.8 cm}|p{1 cm}|p{1.5 cm}|p{0.7 cm}|p{1.2 cm}|}
\begin{tabular}{|>{\centering\arraybackslash}m{0.4 in}||>{\centering\arraybackslash}m{0.69 in}|>{\centering\arraybackslash}m{0.44 in}|>{\centering\arraybackslash}m{0.4 in}|>{\centering\arraybackslash}m{0.39 in}|}
\hline
Category 	& \textbf{\#Candidates flagged} (\textit{distinct})	& Candidates without source & \#Validated  & \textit{\%} of true positives \\ 
\hline
Prodigal 	& \textbf{1504} (\textit{438}) 			& 1487 			& 1253 			&  \textit{97}  \\ \hline

Suicidal 	& \textbf{1495} (\textit{403})			& 1487 			& 1423 			&  \textit{99}  \\ \hline

Greedy 		& \textbf{31,201} (\textit{1524})		& 31,045 		& 1083 			&  \textit{69}  \\ \hline

Total		& \textbf{\totalbuggy} (\textit{\distinctbuggy})	& \withoutsource 	& \validated 	&  \tpweight	\\ \hline 

% Posthumous 	& 853 (\textit{1})			& 853 		& 853 	&  \textit{100}  \\ \hline

\end{tabular}

\caption{Final results using invocation depth~$3$ at block height \texttt{BH}. Column~1 reports number of flagged contracts,
  and the distinct among these. Column~2 shows the number of flagged which have no source code. Column~3 is the subset we sampled
  for concrete validation. Column~4 reports true positive rates; the total here is the average TP rate weighted by the number of validated contracts.}

\label{fig:table}
\end{table}
}}

\subsection{Results}

Table~\ref{fig:table} summarizes the contracts flagged by
\codename. Given the large number of flagged contracts, we select a
random subset for concrete validation, and report on the true positive
rates obtained. We report the number of distinct contracts, calculated
by comparing the hash of the bytecode; however, all percentages are
calculated on the original number of contracts (with duplicates).
\paragraph{Prodigal contracts}
Our tool has flagged $1,504$ candidates contracts ($438$ distinct)
which may leak Ether to an arbitrary Ethereum address, with a true
positive rate of around $97\%$. At block height \texttt{BH}, $46$ of
these contracts hold some Ether.
The concrete validation described in Section~\ref{sec:validation}
succeeds for exploits for $37$ out of $46$ --- these are true
positives, whereas $7$ are false positives. The remaining $2$
contracts leak Ether to an address different from the caller's
address. Note that all of the 37 true positive contracts are alive as
of this writing. For ethical reasons, no exploits were done on
the main blockchain.

Of the remaining $1,458$ contracts which presently do not have Ether
on the public Ethereum blockchain, $24$ have been killed and $42$ have
not been published (as of block height \texttt{BH}). To validate the
remaining alive contracts (in total $1392$) on a private fork, first
we send them Ether from our mining account, and find that $1,183$
contracts can receive Ether.\footnote{These are live and we could
  update them with funds in testing.}  We then concretely
validate whether these contract leak Ether to an arbitrary address.  A
total of $1,156$ out of $1,183$ ($97.72\%$) contracts are confirmed to
be true positives; $27$ ($2.28\%$) are false positives.

%Note, the remaining contracts that do not hold Ether and do not
%accept Ether from our mining account, cannot be confirmed by the
%concrete validation engine, however, some of them can be validated
%individually, at block heights where the contracts are alive and have
%Ether\footnote{We do not proceed with the individual validation as it
%requires significant effort.}.  On the other hand,

For each of the $24$ contracts killed by the block height \texttt{BH},
the concrete validation proceeds as follows. We create a private test
fork of the blockchain, starting from a snapshot at a block height
where the contract is alive.  We send Ether to the contract from one
of our addresses address, and check if the contract leaks Ether to an
arbitrary address. We repeat this procedure for each contract, and
find that all $24$ candidate contracts are true positives.

%In summary, the true positive rate is at most $97.28\%$, the false positive rate is at least $2.72\%$.

\paragraph{Suicidal contracts}
\codename flags $1,495$ contracts ($403$ distinct), including the
\code{ParityWalletLibrary} contract, as found susceptible to being
killed by an arbitrary address, with a nearly $99\%$ true positive
rate.  Out of $1,495$ contracts, $1,398$ are alive at \texttt{BH}.
Our concrete validation engine on a private fork of Ethereum confirm
that $1,385$ contracts (or $99.07\%$) are true positives, \ie, they
can be killed by any arbitrary Ethereum account, while $13$ contracts
(or $0.93\%$) are false positives.  The list of true positives
includes the recent \code{ParityWalletLibrary} contract which was
killed at block height $4,501,969$ by an arbitrary account. Of the
$1,495$ contracts flagged, $25$ have been killed by \texttt{BH}; we
repeat the procedure described previously and cofirmed all of them as
true positives.
%$72$ have not been created by \texttt{BH}.

\paragraph{Greedy contracts}
Our tool flags $31,201$ greedy candidates ($1,524$ distinct), which
amounts to around $3.2\%$ of the contracts present on the blockchain.
The first observation is that \codename deems all but these as
accepting Ether but having states that release them (not locking
indefinitely). To validate a candidate contract as a true positive one
has to show that the contract does not release/send Ether to any
address for any valid trace.  However, concrete validation may not
cover all possible traces, and thus it cannot be used to confirm if a
contract is greedy. Therefore, we take a different strategy and divide
them into two categories:

\begin{enumerate}[label=(\roman*)~]
	
	\item Contracts that accept Ether, but in their bytecode do not have any of the instructions that release Ether 
	(such instructions include  \opn{CALL}, \opn{SUICIDE}, or \opn{DELEGATECALL}). 

      \item Contracts that accept Ether, and in their bytecode have at
        least one of \opn{CALL}, \opn{SUICIDE}
        or \opn{DELEGATECALL}.

\end{enumerate}

\codename flagged $1,058$ distinct contracts from the first
category. We validate that these contracts can receive Ether (we send
Ether to them in a transaction with input data according to the one
provided by the symbolic execution routine). Our experiments show that
$1,057$ out of $1,058$ (\eg, $99.9\%$) can receive Ether and thus are
true positives.  On the other hand, the tool flagged $466$ distinct
contracts from the second category, which are harder to confirm by
testing alone. We resort to manual analysis for a subset of these
which have source code.  Among these, only $25$ have Solidity source
code. With manual inspection we find that none of them are true
positive --- some traces can reach the \texttt{CALL} code, but
\codename failed to reach it in its path exploration. The reasons for
these are mentioned in the Section~\ref{sec:falsepos}.  By
extrapolation (weighted average across $1,083$ validated), we obtain
true positive rate among greedy contracts of $69\%$.

\paragraph{Posthumous Contracts}
Recall that posthumous are contracts that are dead on the blockchain
(have been killed) but still have non-zero Ether balance.  We can find
such contracts by querying the blockchain, \ie, by collecting all
contracts without executable code, but with non-zero balance.  We
found $853$ contracts at a block height of $4,799,998$ that do not
have any compiled code on the blockchain but have positive Ether
balance. Interestingly, among these, $294$ contracts have received
Ether after they became dead.

% \begin{figure}[t]
%   % \fbox{
% % \begin{varwidth}{\dimexpr\textwidth+2\fboxsep-2\fboxrule\relax}
% \centering
%  \begin{subfigure}{0.565\textwidth}
%   \includegraphics[width=.6\linewidth, height=4 cm]{contractstats/sizedisthistlog.png}  
% % \end{varwidth}}
%   \caption{Distribution of contract code size, log scale.}
%   \label{fig:sizedist} 
%  \end{subfigure}
%  \begin{subfigure}{0.565\textwidth}
%    \includegraphics[width=.6\linewidth, height=4 cm]{contractstats/funcdistlog.png}
%    \caption{Distribution of number of functions, log scale.}
%    \label{fig:funcdist} 
%  \end{subfigure}
%  \caption{Contract code characteristics}  
% \label{fig:codedist} 
% \end{figure}

% <<<<<<< HEAD
{
\setlength{\belowcaptionskip}{-7pt}
\setlength{\abovecaptionskip}{-5pt}
{

\begin{figure}[t]
\centering

\begin{lstlisting}[mathescape=true,language=Solidity, frame=none]
bytes20 prev;
function tap(bytes20 nickname) {
    prev = nickname;
    if (prev != nickname) {
      msg.sender.send(this.balance);
    }
}
\end{lstlisting}
\captionof{figure}{A prodigal contract.}
%\caption{\texttt{Buggy} contract.}
\label{fig:leaky} 
%\end{figure*}
\end{figure}    
}}

\vspace{-10pt}

\subsection{Case Studies: True Positives}
Apart from examples presented in section~\ref{sec:bugs}, we now
present true and false postive cases studies. Note that we only
present the contracts with source code for readability. However, the
fraction of flagged contracts with source codes is very low ($1\%$).

\paragraph{Prodigal contracts}
In Figure ~\ref{fig:leaky}, we give an example of a prodigal contract. 
The function \code{tap} seems to lock Ether  because the condition at line $4$, semantically, can never be true. However, the compiler optimization 
of Solidity allows this condition to pass when an input greater than 20 bytes is used to 
call the function \code{tap}. 
Note, on a bytecode level, the EVM can only load chunks of 32 bytes of input data. 
% and decodes it according to the type of argument. 
At line 3 in \code{tap} the first 20 bytes of \code{nickname}  
are assigned to the global variable \code{prev}, while neglecting the remaining 12 bytes. The 
error occurs because EVM at line 4, correctly nullifies the 12 bytes in \code{prev}, but not in \code{nickname}. 
%first to the right and then to the left (to annulate the 12 bytes) it right left shifts  to expand it to 32 bytes before comparing it to 
Thus if \code{nickname} has non-zero values in these 12 bytes then the inequality is true. 
This contract so far has lost $5.0001$ Ether  to different addresses on real Ethereum blockchain.

{
{

\begin{figure}[t]
\centering

\begin{lstlisting}[mathescape=true,language=Solidity, frame=none]
contract Mortal {
  address public owner;
  function mortal() { 
    owner = msg.sender; 
  }
  function kill() { 
    if (msg.sender == owner){
      suicide(owner); 
    }
  }
}
contract Thing is Mortal { /*...*/ }
\end{lstlisting}
\captionof{figure}{The prodigal contract \texttt{Thing}, derived from
  \texttt{Mortal}, leaks Ether to any address by getting killed.}
%\caption{\texttt{Buggy} contract.}
\label{fig:mortal} 
%\end{figure*}
\end{figure}
}} 

A contract may also leak Ether by getting killed
since the semantic of \texttt{SUICIDE} instruction enforce it to send all of its 
balance to an address provided to the instruction. In Figure \ref{fig:mortal}, the
contract \code{Thing}\cite{MortalContract} is inherited from a base
contract \code{Mortal}. The contract implements a review system in which
public reviews an ongoing topic. Among others, the contract has a
\code{kill} function inherited from its base contract which is used to send its balance
to its \code{owner} if its killed. The function \code{mortal}, supposedly a constructor, is misspelled, and thus 
% supposed to be a
%constructor, is not a contructor since the constructor's name must exactly
%match to the contract name.  Hence, the base contract 
anyone can call \code{mortal} to become the \code{owner} of the
contract. Since the derived contract \code{Thing} inherits functions from
contract \code{Mortal}, this vulnerability in the base contract allows an
arbitrary Ethereum account to become the \code{owner} of the derived
contract, to kill it, and to receive its Ether. 

{
{

\begin{figure}[t]
\centering
\begin{lstlisting}[mathescape=true,language=Solidity, frame=none]
function withdraw() public returns (uint) {
  Record storage rec = records[msg.sender];
  uint balance = rec.balance;
  if (balance > 0) {
    rec.balance = 0;
    msg.sender.transfer(balance);
    Withdrawn(now, msg.sender, balance);
  }
  if (now - lastInvestmentTime > 4 weeks) {
    selfdestruct(funder);
  }
  return balance; }
\end{lstlisting}
\captionof{figure}{The \texttt{Dividend} contract can be killed by invoking 
  \texttt{withdraw} if the last investment has been made at
  least 4 weeks ago.}
\label{fig:dividend} 
%\end{figure*}
\end{figure}
}}

\paragraph{Suicidal contracts} A contract can be killed by exploiting an
unprotected \texttt{SUICIDE}  instruction. A trivial example is a public kill function  which
hosts the suicide instruction. Sometimes, \texttt{SUICIDE} is protected
by a weak condition, such as in the contract \code{Dividend} given in Figure~\ref{fig:dividend}. 
This contract allows users to buy shares or withdraw their investment. The
logic of  withdrawing investment is implemented by the \code{withdraw} function.
However, this function has a \code{self_destruct} instruction which can be
executed once the last investment has been made more than $4$ weeks ago.
Hence, if an investor calls this function after $4$ weeks of the last investment,
all the funds  go to the owner of the contract and all the records of
investors are cleared from the blockchain. Though the ether is safe with the owner
, there would be no record of any investment for the owner to return 
 ether to investors.

In the previous example,  one invocation  of \code{withdraw} function was
sufficient to kill the contract. There are, however, contracts which require two
or  more function invocations to be killed. For instance, the contract
\code{Mortal} given in Figure~\ref{fig:mortal} checks whether it is the owner  that calls the 
\code{kill} function. Hence, it requires an attacker to become the owner of
the contract to kill it. So, this contract requires two invocations to be killed: one call to the function \code{mortal} used to become an owner of the
contract and one call to the function \code{kill} to kill the contract. 
A more secure contract would leverage
the \code{mortal} function to a constructor so that the function is called
only once when the contract is deployed. Note, the recent Parity bug similarly also
requires two invocations~\cite{Parity}. 

\paragraph{Greedy contracts} 
The contract \code{SimpleStorage}, given in Figure~\ref{fig:storage},
is an example of a contract that locks Ether indefinitely.  When an
arbitrary address sends Ether along with a transaction invoking the
\code{set} function, the contract balance increases by the amount of
Ether sent.  However, the contract does not have any instruction to
release Ether, and thus locks it on the blockchain.

The \code{payable} keyword has been introduced in Solidity recently to
prevent functions from accepting Ether by default, \ie, a function not
associated with \code{payable} keyword throws if Ether is sent in a
transaction. However, although this contract does not have any
function associated with the \code{payable} keyword, it accepts Ether
since it had been compiled with an older version of Solidity compiler
(with no support for \code{payable}).

% Another way of locking Ether on blockchain is by sending Ether to a dead contract. Although it seems trivial, ethereum 
% framework doesn't warn the users who send Ether to such contracts. Thus, users needs to check if a contract is 
% alive by checkinhg its bytecode value\footnote{Can be done using Json-RPC API provided in go-ethereum} on the blockchain. 
% $288$ contracts received Ether after they were killed on blockchain, at a block height of $4799,998$.
{
{

\begin{figure}[t]
\centering
\begin{lstlisting}[mathescape=true,language=Solidity, frame=none]
contract SimpleStorage {
  uint storedData; address storedAddress;
  event flag(uint val, address addr);

  function set(uint x, address y) {
    storedData = x; storedAddress = y;
  }
  function get() constant
    returns(uint retVal, address retAddr) {
    return (storedData,storedAddress);
  }
}
\end{lstlisting}
\captionof{figure}{A contract that locks Ether.}
\label{fig:storage} 
\end{figure}
}}

%\vspace{-10pt}

\subsection{Case Studies: False Positives}
\label{sec:falsepos} 

%\vspace{-2pt}

We manually analyze cases where \codename's concrete validation fails
to trigger the necessary violation with the produced concrete values,
if source code is available.

% These suggest ways for improving
% \codename in the future.

% \vspace{5pt}

%\vspace{0.1 cm}

\paragraph{Prodigal and Suicidal contracts}
In both of the classes, false positives arise due to two reasons:
\begin{enumerate}[label=(\roman*)~]

\item Our tool performs inter-procedural analysis within a contract,
  but does not transfer control in cross-contract calls. For calls
  from one contract to a function of another contract, \codename
  assigns symbolic variables to the return values. This is imprecise,
  because real executions may only return one value (say
  \texttt{true}) when the call succeeds. 
	
\item \codename may assign values to symbolic variables related to
  block state (\eg, \code{timestamp} and \code{blocknumber}) in cases
  where these values are used to decide the control flow.
  Thus, we may get false positives because those values may be
  different at the concrete validation stage. For instance, in
  Figure~\ref{fig:rng}, the \code{_guess} value depends on the values
  of block parameters, which cannot be forced to take on the concrete
  values found by our analyzer.

\end{enumerate}
{
{

\begin{figure}[t]
\centering
\begin{lstlisting}[mathescape=true,language=Solidity, frame=none]
function confirmTransaction(uint tId)
  ownerExists(msg.sender) {
  confirmations[tId][msg.sender] = true;
  executeTransaction(tId);
}
function executeTransaction(uint tId) {
// In case of majority
  if (isConfirmed(tId)) {
    Transaction tx = transactions[tId];
    tx.executed = true;
    if (tx.destination.call.value(tx.value)&\hfill&(tx.data))
      /*....*/
  }}    
\end{lstlisting}
\captionof{figure}{False positive, flagged as a greedy contract.}
%\caption{\texttt{Buggy} contract.}
\label{fig:multisig} 
%\end{figure*}
\end{figure}
}}

{\setlength{\belowcaptionskip}{-10pt}{

\begin{figure}[t]
\centering
\begin{lstlisting}[mathescape=true,language=Solidity, frame=none]
function RandomNumber() returns(uint) {
  /*....*/
  last = seed^(uint(sha3(block.blockhash(
    block.number),nonces[msg.sender]))*0x000b0007000500030001);
}
function Guess(uint _guess) returns (bool) {
  if (RandomNumber() == _guess) {
    if (!msg.sender.send(this.balance)) throw;
    /*....*/	
  }/*....*/}
\end{lstlisting}
\captionof{figure}{False positive, flagged as a prodigal contract.}
%\caption{\texttt{Buggy} contract.}
\label{fig:rng} 
%\end{figure*}
\end{figure}
}}

%\vspace{0.1 cm}

%\vspace{0.1 cm}

\paragraph{Greedy contracts}
The large share of false positives is attributed to two causes:
\begin{enumerate}[label=(\roman*)~]

\item Detecting a trace which leads to release of Ether may need three
  or more function invocations.  For instance, in
  Figure~\ref{fig:multisig}, the function \code{confirmTransaction}
  has to be executed by the majority of owners for the contract to
  execute the transaction. Our default invocation depth is the reason
  for missing a possible reachable state.

\item Our tool is not able to recover the subtype for the generic
  \code{bytes} type in the EVM semantics.

\item Some contracts release funds only if a random number (usually
  generated using transaction and block parameters) matches a
  predetermined value unlike in the case of the contract in
  Figure~\ref{fig:rng}. In that contract the variable \code{_guess} is
  also a symbolic variable, hence, the solver can find a solution for
  condition on line~$7$. If there is a concrete value in place of
  \code{_guess}, the solver times out since the constraint involves a
  hash function (hard to invert by the SMT solver).

\end{enumerate}

% \subsection{Investigation into Causes} 
% \begin{figure}[t!]
%   % \fbox{
% % \begin{varwidth}{\dimexpr\textwidth+2\fboxsep-2\fboxrule\relax}
% \centering
%  \begin{subfigure}{0.565\textwidth}
%   \includegraphics[width=.4\linewidth]{sizeVsBugsHist/sizevsleakHist.png}  
% % \end{varwidth}}
%   \caption{leak}
%   \label{fig:sizevsleak} 
%  \end{subfigure}
%  \begin{subfigure}{0.565\textwidth}
%    \includegraphics[width=.4\linewidth]{sizeVsBugsHist/sizevslockHist.png}
%    \caption{lock}
%    \label{fig:sizevslock} 
%  \end{subfigure}
%  \begin{subfigure}{0.565\textwidth}
%    \includegraphics[width=.4\linewidth]{sizeVsBugsHist/sizevskillHist.png}
%    \caption{lock}
%    \label{fig:sizevskill} 
%  \end{subfigure}
%  \caption{Variation of bugs with size of contracts}  
% \label{fig:sizevsbugsdist} 
% \end{figure}

\vspace{-10pt}

\subsection{Summary and Observations}

The symbolic execution engine of \codename flags \totalbuggy
contracts. % with an invocation depth of $2$.
With concrete validation engine or manual inspection, we have
confirmed that around $97\%$ of prodigal, $97\%$ of suicidal and
$69\%$ of greedy contracts are true positive.  The importance of
analyzing the bytecode of the contracts, rather than Solidity source
code, is demonstrated by the fact that only $1\%$ of all contracts
have source code. Further, among all flagged contracts, only
\withsource have verified source codes according to the widely used
platform Etherscan, or in percentages only $1.06\%$, $0.47\%$ and
$0.49\%$, in the three categories of prodigal, suicidal, and greedy,
respectively.  We refer the reader to Table~\ref{fig:table} for the
exact summary of these results.

Furthermore, the maximal amount of Ether that could have been
withdrawn from prodigal and suicidal contracts, before the block
height \texttt{BH}, is nearly $4,905$ Ether, or $5.9$ million US
dollars\footnote{Calculated at
  $1,210$ USD/Eth~\cite{EtherscanWebsite}.}
according to the exchange rate at the time of this writing.  In
addition, $6,239$ Ether ($7.5$ million US dollars) is locked inside
posthumous contracts currently on the blockchain, of which $313$ Ether
($379,940$ US dollars) have been sent to dead contracts after they
have been killed.

%\paragraph{Observation} 

Finally, the analysis given in Table~\ref{fig:table2} shows the number
of flagged contracts for different invocation depths from $1$ to
$4$. We tested $25,000$ contracts being for greedy, and $100,000$ for
remaining categories, inferring that increasing depth improves results
marginally, and an invocation depth of $3$ is an optimal tradeoff
point.

{\setlength{\belowcaptionskip}{-10pt}{
\begin{table}[t]
\centering
\footnotesize
\begin{tabular}{|>{\centering\arraybackslash}m{0.55 in}||>{\centering\arraybackslash}m{0.55 in}|>{\centering\arraybackslash}p{0.55 in}|>{\centering\arraybackslash}m{0.55 in}|}
\hline
Inv. depth & Prodigal & Suicidal & Greedy \\
\hline
1 & 131 & 127 & 682  \\
\hline
2 & 156 & 141 & 682 \\
\hline
3 & 157 & 141 & 682 \\
\hline
4 & 157 & 141 & 682 \\
\hline
\end{tabular}
\caption{The table shows number of contracts flagged
for various invocation depths. This analysis is done on a 
random subset of $25,000$--$100,000$ contracts. 
% The table shows an increase of just $10\%$ in the number 
% of contracts flagged from invocation depth $1$ to $2$ and 
% almost no change in that number as depth increases further. 
%
%The table suggests that after an invocation depth of $4$ the number
%of candidates flagged, might almost remain constant.
}
\label{fig:table2}
\end{table}
}}

\vspace{-10pt}

\section{Related Work}
\label{sec:related}

%\vspace{-5pt}

% Security and safety properties of smart contracts have received a lot
% of attention since a number of costly bugs and exploits took
% place~\cite{dao,parity-bug}.

\paragraph{Dichotomy of smart contract bugs}
\label{sec:dich-smart-contr}
%
% The majority of the bugs in Ethereum-style smart contracts are due to
% the de-facto high-level implementation language,
% Solidity~\cite{Solidity}, whose runtime behaviour that diverge from
% the ``intuitive understanding'' of the language by the developers.
%
The early work by Delmolino~\etal~\cite{Delmolino-al:FC16}
distinguishes the following classes of problems: (a) contracts that do
not refund their users, (b) missing encryptions of sensitive user data
and (c) lack of incentives for the users to take certain actions. The
property (a) is the closest to our notion of \emph{greedy}. While that
outlines the problem and demonstrates it on series of simple examples
taught in a class, they do not provide a systematic approach for
detection of smart contracts prone to this issue.
Later works on contract safety and security identify potential bugs,
related to the concurrent transactional
executions~\cite{Sergey-Hobor:WTSC17}, mishandled
exceptions~\cite{Luu-al:CCS16}, overly extensive gas
consumption~\cite{Chen-al:SANER17} and implementations of fraudulent
financial schemes~\cite{Bartoletti-al:Arxiv17}.\footnote{See the
  works~\cite{Atzei-al:POST17,consensys-practices} for a survey of
  known contract issues.}

In contrast to all those work, which focus on bad implementation
practices or misused language semantics, we believe, our
characterisation of several classes of contract bugs, such as greedy,
prodigal, \etc, is novel, as they are stated in terms of properties
execution traces rather than particular instructions taken/states
reached.

\vspace{15pt}

\paragraph{Reasoning about smart contracts}
%
% Several tools have been proposed to automatic detection of
% vulnirabilities in smart contracts, as well as for formal contract
% verification.
%
\tname{Oyente}~\cite{Luu-al:CCS16,Oyente} was the first symbolic
execution-based tool that provided analysis targeting several specific
issues: (a) mishandled exceptions, (b) transaction-ordering
dependence, (c)~timestamp dependence and (d)
reentrancy~\cite{reentrancy}, thus remedying the corner cases of
Solidity/EVM semantics (a) as well as some programming anti-patterns
(b)--(d).

Other tools for symbolic analysis of EVM and/or EVM have been
developed more recently: \tname{Manticore}~\cite{manticore},
\tname{Mythrill}~\cite{mythril,mueller-z3},
\tname{Securify}~\cite{securify}, and
\tname{KEVM}~\cite{Hildenbrandt:KEVM17,erc20-k}, all focusing on
detecting \emph{low-level} safety violations and vulnerabilities, such
as integer overflows, reentrancy, and unhandled exceptions, \etc,
neither of them requiring reasoning about contract execution traces.
%
% While it does not seem impossible to extend all these frameworks for
% handling trace-based properties discussed in this work, this has not
% been done yet, thus we cannot conduct a formal comparison. 
%
A very recent work by Grossman~\etal~\cite{Grossman-al:POPL18} similar
to our in spirit and providing a dynamic analysis of execution traces,
focuses exclusively on detecting \emph{non-callback-free} contracts
(\ie, prone to reentrancy attacks)---a vulnerability that is by now
well studied.

Concurrently with our work, Kalra~\etal developed
\tname{Zeus}~\cite{Kalra-al:NDSS18}, a framework for automated
verification of smart contracts using abstract interpretation and
symbolic model checking, accepting user-provided \emph{policies} to
verify for.
Unlike \codename, \tname{Zeus} conducts policy checking at a level of
LLVM-like intermediate representation of a contract, obtained from
Solidity code, and leverages a suite of standard tools, such as
off-the-shelf constraint and SMT
solvers~\cite{Gurfinkel-al:CAV15,McMillan:VMCAI07,deMoura-Bjorner:TACAS08}.
\tname{Zeus} does not provide a general framework for checking trace
properties, or under-approximating liveness properties.

Various versions of EVM semantics~\cite{Gavin-al:yellow-paper} were
implemented in Coq~\cite{Hirai:EVMCoq},
Isabelle/HOL~\cite{Hirai:WTSC17,Amani-al:CPP18},
$\text{F}^{\star}$~\cite{Bhargavan-al:PLAS16},
Idris~\cite{ethereum-idris:16}, and
Why3~\cite{eth-why3,Filliatre-Paskevich:ESOP13}, followed by
subsequent mechanised contract verification efforts. However, none of
those efforts considered trace properties in the spirit of what we
defined in Section~\ref{sec:properties}.

%\paragraph{Towards verification-friendly contract languages}
%
% Analysing for trace properties of interest at the level of EVM
% bytecode leads to the loss of precision and the need to take
% compromises in a tool implementation, for the sake of performance,
% leading to both missed bugs and fals positives. 
% %
% A more principled language model, where global actions, such as
% sending Ether or terminating a contract, would be treated differently
% from instructions performing ordinary computations, would make such an
% analysis more tractable and scalable. 

Several contract languages were proposed recently that distinguish
between \emph{global actions} (\eg, sending Ether or terminating a
contract) and \emph{instructions} for ordinary
computations~\cite{bamboo,scilla}, for the sake of simplified
reasoning about contract executions.
For instance, the work on the contract
language~\tname{Scilla}~\cite{scilla} shows how to encode in
Coq~\cite{Coq-manual} and formally prove a property, which is very
similar to a contract being non-\emph{leaky}, as per
Definition~\ref{def:leaky} instantiated with a non-trivial side
condition~$R$.

\vspace{-20pt}

\section{Conclusion}
\label{sec:conclusion}

\vspace{-5pt}

We characterize vulnerabilities in smart contracts that are checkable
as properties of an entire execution trace (possibly infinite sequence
of their invocations). We show three examples of such trace
vulnerabilities, leading to greedy, prodigal and suicidal contracts.
%We build a symbolic analysis tool to find these.
Analyzing $970,898$ contracts, our new tool \codename flags thousands
of contracts vulnerable at a high true positive rate.

% Our new tool \codename performs symbolic execution of contract
% bytecode directly. At a scale of nearly one million contracts,
% \codename flags thousands of contracts as vulnerable, and successfully
% generates exploits for $69$--$99\%$ of the subset we sample for
% validation.
% %>>>>>>> 54e44f79e874e8c4b331b47f5ce1075b6b6f5e07

% \paragraph{Aknowledgements}
% %
% We thank 
% %
% Andreea Costea,
% Teodora B\u{a}lu\cb{t}\u{a},
% Shiqi Shen and
% Muoi Tran
% %
% for their comments on earlier drafts of this paper.
% %
% Sergey's research was supported by EPSRC First Grant EP/P009271/1
% ``Program Logics for Compositional Specification and Verification of
% Distributed Systems''.

%\input{discussion}

%\input{overview}
%\input{issues}
% \input{language}

%% \section{Appendix Title}
%%
%% This is the text of the appendix, if you need one.
%%
%% \acks
%%
%% Acknowledgments, if needed.

%\setlength{\bibsep}{1.8pt}  
{\footnotesize 
\bibliographystyle{IEEEtran}
\bibliography{references,proceedings}

\begin{thebibliography}{10}
\providecommand{\url}[1]{#1}
\csname url@samestyle\endcsname
\providecommand{\newblock}{\relax}
\providecommand{\bibinfo}[2]{#2}
\providecommand{\BIBentrySTDinterwordspacing}{\spaceskip=0pt\relax}
\providecommand{\BIBentryALTinterwordstretchfactor}{4}
\providecommand{\BIBentryALTinterwordspacing}{\spaceskip=\fontdimen2\font plus
\BIBentryALTinterwordstretchfactor\fontdimen3\font minus
  \fontdimen4\font\relax}
\providecommand{\BIBforeignlanguage}[2]{{%
\expandafter\ifx\csname l@#1\endcsname\relax
\typeout{** WARNING: IEEEtran.bst: No hyphenation pattern has been}%
\typeout{** loaded for the language `#1'. Using the pattern for}%
\typeout{** the default language instead.}%
\else
\language=\csname l@#1\endcsname
\fi
#2}}
\providecommand{\BIBdecl}{\relax}
\BIBdecl

\bibitem{Parity}
\BIBentryALTinterwordspacing
A.~Akentiev, ``Parity multisig github.'' [Online]. Available:
  \url{https://github.com/paritytech/parity/issues/6995}
\BIBentrySTDinterwordspacing

\bibitem{Luu-al:CCS16}
L.~Luu, D.~Chu, H.~Olickel, P.~Saxena, and A.~Hobor, ``Making smart contracts
  smarter,'' in \emph{CCS}.\hskip 1em plus 0.5em minus 0.4em\relax {ACM}, 2016,
  pp. 254--269.

\bibitem{Oyente}
\BIBentryALTinterwordspacing
``{Oyente: An Analysis Tool for Smart Contracts},'' 2018. [Online]. Available:
  \url{https://github.com/melonproject/oyente}
\BIBentrySTDinterwordspacing

\bibitem{Kalra-al:NDSS18}
S.~Kalra, S.~Goel, M.~Dhawan, and S.~Sharma, ``Zeus: Analyzing safety of smart
  contracts,'' in \emph{NDSS}, 2018, to appear.

\bibitem{securify}
\BIBentryALTinterwordspacing
``{Securify: Formal Verification of Ethereum Smart Contracts},'' 2018.
  [Online]. Available: \url{http://securify.ch/}
\BIBentrySTDinterwordspacing

\bibitem{dao}
M.~{del Castillo}, ``{The DAO Attacked: Code Issue Leads to \$60 Million Ether
  Theft},'' June 17, 2016.

\bibitem{Governmental}
\BIBentryALTinterwordspacing
``Governmental's $1100 eth$ jackpot payout is stuck because it uses too much
  gas.'' [Online]. Available:
  \url{https://www.reddit.com/r/ethereum/comments/4ghzhv/}
\BIBentrySTDinterwordspacing

\bibitem{Gavin-al:yellow-paper}
\BIBentryALTinterwordspacing
G.~Wood, ``Ethereum: A secure decentralised generalised transaction ledger.''
  [Online]. Available: \url{https://ethereum.github.io/yellowpaper/paper.pdf}
\BIBentrySTDinterwordspacing

\bibitem{Solidity}
\BIBentryALTinterwordspacing
\emph{{Solidity: High-Level Language for Implementing Smart Contracts}}.
  [Online]. Available: \url{http://solidity.readthedocs.io/}
\BIBentrySTDinterwordspacing

\bibitem{Nakamoto:08}
\BIBentryALTinterwordspacing
S.~Nakamoto, ``Bitcoin: A peer-to-peer electronic cash system,'' 2008.
  [Online]. Available: \url{http://bitcoin.org/bitcoin.pdf}
\BIBentrySTDinterwordspacing

\bibitem{Pirlea-Sergey:CPP18}
G.~P{\^{\i}}rlea and I.~Sergey, ``Mechanising blockchain consensus,'' in
  \emph{CPP}.\hskip 1em plus 0.5em minus 0.4em\relax ACM, 2018, pp. 78--90.

\bibitem{parity-bug}
J.~Alois, ``{Ethereum Parity Hack May Impact ETH 500,000 or \$146 Million},''
  2017.

\bibitem{Parityinnocent}
\BIBentryALTinterwordspacing
``The guy who blew up parity didn't know what he was doing.'' [Online].
  Available: \url{https://www.reddit.com/r/CryptoCurrency/comments/7beos3/}
\BIBentrySTDinterwordspacing

\bibitem{Chen-al:SANER17}
T.~Chen, X.~Li, X.~Luo, and X.~Zhang, ``Under-optimized smart contracts devour
  your money,'' in \emph{{IEEE} 24th International Conference on Software
  Analysis, Evolution and Reengineering, {SANER}}, 2017, pp. 442--446.

\bibitem{mythril}
\BIBentryALTinterwordspacing
``Mythril,'' 2018. [Online]. Available:
  \url{https://github.com/b-mueller/mythril/}
\BIBentrySTDinterwordspacing

\bibitem{mueller-z3}
\BIBentryALTinterwordspacing
B.~Mueller, ``{How Formal Verification Can Ensure Flawless Smart Contracts},''
  January 2018. [Online]. Available: \url{https://goo.gl/9wUFE1}
\BIBentrySTDinterwordspacing

\bibitem{manticore}
\BIBentryALTinterwordspacing
``{Manticore},'' 2018. [Online]. Available:
  \url{https://github.com/trailofbits/manticore}
\BIBentrySTDinterwordspacing

\bibitem{Godefroid:2011}
P.~Godefroid, ``Higher-order test generation,'' in \emph{Proceedings of the
  32Nd ACM SIGPLAN Conference on Programming Language Design and
  Implementation}, ser. PLDI '11, 2011.

\bibitem{deMoura-Bjorner:TACAS08}
L.~M. {de Moura} and N.~Bj{\o}rner, ``{Z3:} an efficient {SMT} solver,'' in
  \emph{TACAS}, ser. LNCS, vol. 4963.\hskip 1em plus 0.5em minus 0.4em\relax
  Springer, 2008, pp. 337--340.

\bibitem{goEthereum}
\BIBentryALTinterwordspacing
``Go-ethereum.'' [Online]. Available:
  \url{https://github.com/ethereum/go-ethereum}
\BIBentrySTDinterwordspacing

\bibitem{etherscanSourceCodes}
\BIBentryALTinterwordspacing
``Etherscan verified source codes.'' [Online]. Available:
  \url{https://etherscan.io/contractsVerified}
\BIBentrySTDinterwordspacing

\bibitem{MortalContract}
\BIBentryALTinterwordspacing
``Contract mortal.'' [Online]. Available:
  \url{https://etherscan.io/address/0x4671ebe586199456ca28ac050cc9473cbac829eb#code}
\BIBentrySTDinterwordspacing

\bibitem{EtherscanWebsite}
\BIBentryALTinterwordspacing
``Etherscan.'' [Online]. Available: \url{https://etherscan.io/}
\BIBentrySTDinterwordspacing

\bibitem{Delmolino-al:FC16}
K.~Delmolino, M.~Arnett, A.~E. Kosba, A.~Miller, and E.~Shi, ``Step by step
  towards creating a safe smart contract: Lessons and insights from a
  cryptocurrency lab,'' in \emph{{FC} 2016 International Workshops}, ser. LNCS,
  vol. 9604.\hskip 1em plus 0.5em minus 0.4em\relax Springer, 2016, pp. 79--94.

\bibitem{Sergey-Hobor:WTSC17}
I.~Sergey and A.~Hobor, ``{A Concurrent Perspective on Smart Contracts},'' in
  \emph{1st Workshop on Trusted Smart Contracts}, ser. LNCS, vol. 10323.\hskip
  1em plus 0.5em minus 0.4em\relax Springer, 2017, pp. 478--493.

\bibitem{Bartoletti-al:Arxiv17}
M.~Bartoletti, S.~Carta, T.~Cimoli, and R.~Saia, ``Dissecting ponzi schemes on
  ethereum: identification, analysis, and impact,'' \emph{CoRR}, vol.
  abs/1703.03779, 2017.

\bibitem{Atzei-al:POST17}
N.~Atzei, M.~Bartoletti, and T.~Cimoli, ``{A Survey of Attacks on Ethereum
  Smart Contracts (SoK)},'' in \emph{POST}, ser. LNCS, vol. 10204.\hskip 1em
  plus 0.5em minus 0.4em\relax Springer, 2017, pp. 164--186.

\bibitem{consensys-practices}
\BIBentryALTinterwordspacing
{ConsenSys Diligence}, ``{Ethereum Smart Contract Security Best Practices},''
  2018. [Online]. Available:
  \url{https://consensys.github.io/smart-contract-best-practices}
\BIBentrySTDinterwordspacing

\bibitem{reentrancy}
\BIBentryALTinterwordspacing
E.~G. Sirer, ``{Reentrancy Woes in Smart Contracts}.'' [Online]. Available:
  \url{http://hackingdistributed.com/2016/07/13/reentrancy-woes/}
\BIBentrySTDinterwordspacing

\bibitem{Hildenbrandt:KEVM17}
E.~Hildenbrandt, M.~Saxena, X.~Zhu, N.~Rodrigues, P.~Daian, D.~Guth, and
  G.~Rosu, ``{KEVM: A Complete Semantics of the Ethereum Virtual Machine},''
  Tech. Rep., 2017.

\bibitem{erc20-k}
\BIBentryALTinterwordspacing
G.~Rosu, ``{ERC20-K: Formal Executable Specification of ERC20},'' December
  2017. [Online]. Available: \url{https://runtimeverification.com/blog/?p=496}
\BIBentrySTDinterwordspacing

\bibitem{Grossman-al:POPL18}
S.~Grossman, I.~Abraham, G.~Golan{-}Gueta, Y.~Michalevsky, N.~Rinetzky,
  M.~Sagiv, and Y.~Zohar, ``Online detection of effectively callback free
  objects with applications to smart contracts,'' \emph{{PACMPL}}, vol.~2, no.
  {POPL}, pp. 48:1--48:28, 2018.

\bibitem{Gurfinkel-al:CAV15}
A.~Gurfinkel, T.~Kahsai, A.~Komuravelli, and J.~A. Navas, ``{The SeaHorn
  Verification Framework},'' in \emph{CAV, Part {I}}, ser. LNCS, vol.
  9206.\hskip 1em plus 0.5em minus 0.4em\relax Springer, 2015, pp. 343--361.

\bibitem{McMillan:VMCAI07}
K.~L. McMillan, ``{Interpolants and Symbolic Model Checking},'' in
  \emph{VMCAI}, ser. LNCS, vol. 4349.\hskip 1em plus 0.5em minus 0.4em\relax
  Springer, 2007, pp. 89--90.

\bibitem{Hirai:EVMCoq}
\BIBentryALTinterwordspacing
Y.~Hirai, ``{Ethereum Virtual Machine for Coq (v0.0.2)},'' Published online on
  5 March 2017. [Online]. Available: \url{https://goo.gl/DxYFwK}
\BIBentrySTDinterwordspacing

\bibitem{Hirai:WTSC17}
------, ``{Defining the Ethereum Virtual Machine for Interactive Theorem
  Provers},'' in \emph{1st Workshop on Trusted Smart Contracts}, ser. LNCS,
  vol. 10323.\hskip 1em plus 0.5em minus 0.4em\relax Springer, 2017, pp.
  520--535.

\bibitem{Amani-al:CPP18}
S.~Amani, M.~B\'{e}gel, M.~Bortin, and M.~Staples, ``{Towards Verifying
  Ethereum Smart Contract Bytecode in Isabelle/HOL},'' in \emph{CPP}.\hskip 1em
  plus 0.5em minus 0.4em\relax ACM, 2018, pp. 66--77.

\bibitem{Bhargavan-al:PLAS16}
K.~Bhargavan, A.~Delignat-Lavaud, C.~Fournet, A.~Gollamudi, G.~Gonthier,
  N.~Kobeissi, N.~Kulatova, A.~Rastogi, T.~Sibut-Pinote, N.~Swamy, and
  S.~Zanella-B{\'e}guelin, ``Formal verification of smart contracts: Short
  paper,'' in \emph{PLAS}.\hskip 1em plus 0.5em minus 0.4em\relax ACM, 2016,
  pp. 91--96.

\bibitem{ethereum-idris:16}
J.~Pettersson and R.~Edstr\"{o}m, ``{Safer Smart Contracts through Type-Driven
  Development},'' Master's thesis, Chalmers University of Technology, Sweden,
  2016.

\bibitem{eth-why3}
\BIBentryALTinterwordspacing
C.~Reitwiessner, ``Formal verification for solidity contracts,'' 2015.
  [Online]. Available:
  \url{https://forum.ethereum.org/discussion/3779/formal-verification-for-solidity-contracts}
\BIBentrySTDinterwordspacing

\bibitem{Filliatre-Paskevich:ESOP13}
J.~Filli{\^{a}}tre and A.~Paskevich, ``{Why3 - Where Programs Meet Provers},''
  in \emph{ESOP}, ser. LNCS, vol. 7792.\hskip 1em plus 0.5em minus 0.4em\relax
  Springer, 2013, pp. 125--128.

\bibitem{bamboo}
\BIBentryALTinterwordspacing
``{Bamboo},'' 2018. [Online]. Available:
  \url{https://github.com/pirapira/bamboo}
\BIBentrySTDinterwordspacing

\bibitem{scilla}
I.~Sergey, A.~Kumar, and A.~Hobor, ``Scilla: a smart contract
  intermediate-level language,'' \emph{CoRR}, vol. abs/1801.00687, 2018.

\bibitem{Coq-manual}
\BIBentryALTinterwordspacing
{Coq Development Team}, \emph{{The Coq Proof Assistant Reference Manual -
  Version 8.7}}, 2018. [Online]. Available: \url{http://coq.inria.fr/}
\BIBentrySTDinterwordspacing

\end{thebibliography}
}

\end{document}